\documentclass[final, 3p, 12pt, times, onecolumn]{elsarticle}

\usepackage{amssymb}
\usepackage{amsmath}
\usepackage{amstext}
\usepackage{subcaption}
\usepackage{graphicx}
\usepackage{dcolumn}
\usepackage[usenames,dvipsnames,svgnames,table]{xcolor}
\usepackage[colorlinks=true,
            linkcolor=green,
            urlcolor=cyan,
            citecolor=green]{hyperref}
\usepackage{bm}
\usepackage{enumerate}
\usepackage{lipsum}
\usepackage{epstopdf}
\usepackage{float}
\usepackage{tabularx}
\usepackage{color}
\usepackage{longtable}
\usepackage{rotating}
\usepackage{multicol,lipsum}
\usepackage{cuted}
\usepackage{flushend}
\biboptions{sort&compress}

\journal{Atoms}

\definecolor{mycolor}{rgb}{0.0, 0.5, 0.7}

\makeatletter
\AtBeginDocument{\def\@citecolor{mycolor}
                 \def\@linkcolor{mycolor}
                 \def\@urlcolor{mycolor}}
\makeatother

\begin{document}
\begin{frontmatter}



\title{[Transitional strength under plasma] Precise estimations of  astrophysically relevant electromagnetic transitions of Ar$^{7+}$, Kr$^{7+}$, Xe$^{7+}$, and Rn$^{7+}$  under plasma atmosphere}
\author[Biswas, Bowmik  et al.] {Swapan Biswas$^{1}$,
Anal Bhowmik$^{2,3}$, Arghya Das$^4$, Radha Raman Pal$^1$, Sonjoy Majumder$^4$
\\${^1}$Department of Physics, Vidyasagar University, Midnapore-721102, West Bengal, India.\\
$^2$ Department of Physics, University of Haifa, Haifa 3498838, Israel.\\
$^3$ Haifa Research Center for Theoretical Physics and Astrophysics, University of Haifa, Haifa 3498838, Israel.\\
${^4}$Department of Physics, Indian Institute of Technology Kharagpur, Kharagpur-721302, India.}

	\begin{abstract}
		The growing interest in atomic structures of moderately-stripped alkali-like ions in diagnostic study and modeling of astrophysical and laboratory plasma makes an accurate many-body study of atomic properties inevitable. This work presents transition line parameters in the absence or presence of plasma atmosphere for astrophysically important candidates, Ar$^{7+}$, Kr$^{7+}$, Xe$^{7+}$, and Rn$^{7+}$. We employ relativistic coupled-cluster (RCC) theory, a well-known correlation exhaustive method. In the case of a plasma environment, we use Debye Model. Our calculations agree with experiments available in the literature for ionization potentials, transition strengths of allowed and forbidden selections, and lifetimes of several low-lying states. The unit ratios of length and velocity forms of transition matrix elements are the critical estimation of the accuracy of the transition data presented here, especially for a few presented first time in the literature. We do compare our findings with the available recent theoretical results. Our reported data can be helpful to the astronomer in estimating the density of the plasma environment around the astronomical objects or in the discovery of observational spectra corrected by that environment. The present results should be advantageous in the modeling and diagnostics laboratory plasma, whereas the calculated ionisation potential depression parameters reveal important characteristics of atomic structure.
	\end{abstract}
 \begin{highlights}
\item Atomic spectroscopy for Ar$^{7+}$, Kr$^{7+}$, Xe$^{7+}$ and Rn$^{7+}$ ions with high accuracy
\item  Plasma screened ionization potential, atomic transition amplitudes and rates
\item  ionisation potential depression parameters 
\end{highlights}
	
\end{frontmatter}
	
	\section{Introduction}
	Barlow et al. \cite{Barlow2013} first observed noble gas molecules in the interstellar medium. The other detections of noble gas elements,  either in diatomic \cite{Gusten2019, Benna2015, Muller2015, Schilke2014} or ionic forms \cite{Taresch1997} in space at UV and IR spectrum, motivate further observations of these species in the universe. It is well known that the atomic and spectroscopic processes are valuable diagnostics for plasma atmosphere in the laboratory or Astronomy. Noble gas atoms are known to be chemically inactive and require high energy to ionize. But, once ionized, their reaction rates are rather fast. Over the years, spectroscopic properties of ionized noble gas atoms have become popular, and observers have started to detect them in space \cite{Enrique2007}. On the other hand, alkali-like ions have emerged as the standard test beds for detailed investigation of current relativistic atomic calculations due to their adequately simple but highly correlated electronic structures \cite{Panigrahi1989, Owusu1977, Xiao-Li2007, Roy1976, Dutta2013}. Accurate theoretical and experimental determinations of the transition line parameters and excited-state lifetimes of highly stripped ions are collaborative with the astronomer to investigate dynamics, chemical compositions, opacity, density, and temperature distributions of the distant galaxy \cite{Herrnstein1999}, planetary nebulae, and even entire the interstellar medium \cite{Dimitrijevc2003, Elden1942, Eidelsberg1981, Mcwhirter1984, Elden1984,  Sterling2002, O'Toole2004, Chayer2005, Vennes2005, Curtis1989, Dimitrijevc2015}. Furthermore, one requires the accurate atomic data of different isotopes of noble gas elements to understand the production of heavy elements in the stellar medium by radiative r- and s-processes  \cite{Beer1983, Clayton1978}. The data of energy spectra of moderate to high-stripped ions are required for precise astrophysical and laboratory plasma modelling. All these physical facts and figures motivate us to investigate the transition line parameters and lifetimes of septuple ionized  astrophysically pertinent inert gases, such as Ar$^{7+}$, Kr$^{7+}$, Xe$^{7+}$,  and Rn$^{7+}$.

	In the series, Ar$^{7+}$ is well-studied in literature.  Berry et al. \cite{Berry1974}  observed 74 lines of Ar$^{7+}$ using the beam-foil technique.  In 1982, Striganov and Odintsova \cite{Striganov1982} published the observed lines of  Ar$^{+}$ through Ar$^{8+}$. \cite{Jupen1990, Biemont1999, Fischer2006} applied the multi-configuration Dirac-Fock (MCDF) method to calculate the autoionization spectrum,  energy levels, transition rates, oscillator strengths, and lifetimes of Ar$^{7+}$. 
	Saloman \cite{Saloman2010} identified the energy spectra of Ar$^{+}$ to Ar$^{17+}$, which he studied from the year 2006 to 2009
	employing beam foil Spectroscopy (BFS), electron beam ion trap (EBIT), laser-excited plasmas, fusion devices, astronomical observations, and {\it ab initio} calculations with quantum electrodynamic corrections.

	Similarly,  krypton ion spectra were  detected in the interstellar medium \cite{Cardelli1997, Cardelli1991},  the galactic disc \cite{Cartledge2003}, and the planetary nebulae \cite{Dinerstein2001}. Fine structure intervals, fine structure inversions, and core-polarization study of the Kr$^{7+}$ ion were performed by  different groups \cite{Cheng1978, Reader1991, Boduch1992} including third-order many-body perturbation theory and M\o{}ller-Pleset perturbation theory \cite{Johnson1990, Vilkas2000}  for the energy levels.  
	
	It is found that  Cu I isoelectronic sequence ions are prominent impurities at high-temperature magnetically confined plasmas \cite{Hinnov1976}, and their emission spectra are  observed under the spark sources \cite{Alexander1971, Reader1977, Curtis1977, Joshi1978}  of the laser-produced plasma \cite{Reader1979, Mansfield1979}  and in the beam-foil excitations \cite{Druetta1976, Irwin1976, Knystautas1977}. The  abundance of photospheric lines of trans-iron group elements in the emission spectra of the white dwarfs opens a new way of studying their radiative transfer mechanism \cite{Werner2012}. The presence of the spectral lines of Cu-like ions motivates more accurate determination of atomic data of the radiative properties of these ions for modelling the chemical abundances. These studies are essential for deducing the stellar parameters  necessary to investigate the environmental condition of the white dwarfs. There are studies of electronic properties for Xe$^{7+}$ using various  many-Body methods \cite{Biemont2007, Migdalek2000, Safronova2003, Glowacki2009, Ivanova2011}. Dimitrijevc et al. \cite{Dimitrijevc2015} identified the importance of Stark broadening at the spectral lines observed in extremely metal-poor halo PNH4-1 in primordial supernova \cite{Otsuka2013}. However, we study Ar$^{+7}$, Kr$^{7+}$, and Xe$^{7+}$ here again to mitigate the lack of all-order many-body calculations or precise experiments and to estimate their spectroscopic properties under a plasma environment. Recent past, one of the present authors \cite{Bhowmik2017a} studied Xe$^{7+}$ exclusively as a single valance system without the plasma screening effect. 
	
	Unlike other noble gas ions, studies of radon ions are rare. However, there are a few many-body calculations on Rn$^{+}$ \cite{Pernpointner2012}, and Rn$^{2+}$ \cite{Eser2018}. The observation of several forbidden lines of Kr and Xe ions in the planetary nebula NGC 7027 was reported recent past \cite{Pequignot1994}. For  Rn$^{7+}$, only Migdalek \cite{Migdalek2018} computed a few energy levels and oscillator strengths of allowed transitions  using the Dirac-Fock method corrected by the core-polarisation effect. 
	
	The aim of the paper is to estimate (a) energies of the ground and low-lying excited states, (b) the oscillator strengths of electromagnetically allowed transitions, (c) transition probabilities of the forbidden transitions, and (d) lifetimes for a few excited states of Ar$^{7+}$, Kr$^{7+}$, Xe$^{7+}$, and Rn$^{7+}$ using the relativistic coupled-cluster (RCC) method \cite{Lindgren1987, Bishope1991, Roy2015}. The accuracies of the RCC calculations are well-established by our group for different applications \cite{Dutta2011, Dutta2012, Dutta2016,  Bhowmik2017b, Bhowmik2018, Bhowmik2020, Bhowmik2022, Biswas2018, Das2020, Das2022}. Our special effort here is to study the plasma screening effect on the radiative transition parameters. It is obvious that the nuclear attraction to the bound electrons of atoms or ions immersed in plasma is screened by the neighbouring ions and the free electrons. The essential feature to note is that the electron correlation of atomic systems in this environment is remarkably different from their corresponding isolated candidate.  Therefore, the screening estimations on the transition parameters play a crucial role in the precise diagnostics of plasma temperature and density in the emitting region. In the plasma environment, the ionization potentials decrease gradually with the increasing strength of plasma screening  \cite{Mondal2013} till they become zero at some critical parameter of plasma. Beyond these critical values of plasma, the states become a continuum state. The corresponding ionization potential beyond which instability occurs is known as ionisation potential depression (IPD) according to the Stewart–Pyatt (SP) model \cite{Stewart1966}. Accurate determination of the IPD can infer much useful information about the plasma atmosphere, such as providing the proper equation of the state, estimating radiate opacity of stellar plasma, and internal confinement fusion plasma etc. We have investigated the change in spectroscopic properties of  Ar$^{7+}$, Kr$^{7+}$, Xe$^{7+}$, and Rn$^{7+}$ in the plasma environment. 
	
		\section{Theory}
	Precise generation of wave functions is important for accurately estimating the atomic properties of few-electron monovalent ions presented in this paper. Here we employ a non-linear version of the well-known RCC theory, a many-body approach which exhaustively pools together correlations. Initially, we solve the Dirac-Coulomb Hamiltonian $H$ satisfying eigenvalue equation $H|\Phi\rangle=E_0|\Phi\rangle$ to generate closed-shell atomic wave function under the potential of $(N-1)$ electrons where
	$$ H=\sum_{i}\left(c{\bf \alpha}_{i}\cdot{\bf p}_{i}+(\beta_{i}-1) c^{2}+V_{\mathrm{nuc}}({\bf r}_{i})+\sum_{j<i}\left( \frac{1}{{\bf r}_{ij}}\right)\right).$$
	Here standard notations are used for all the variables. A single valence reference state for the RCC calculation is generated by adding a single electron in the $\text{v}$-th orbital following Koopman's theorem \cite{Szabo1996}. In RCC formalism, the single valence correlated state $|\Psi_\text{v}\rangle$ is connected with single valence reference state $|\Phi_\text{v}\rangle$ as
	\begin{equation}
	|\Psi_\text{v}\rangle = e^T\{1+S_\text{v}\}|\Phi_\text{v}\rangle,
	\hspace {0.2cm} \text{where}\ |\Phi_\text{v}\rangle = a^{\dagger}_\text{v}|\Phi\rangle.   
	\end{equation}                  
	The operator $T$ deals with the excitations from core orbitals and can generate core-excited configurations from closed-shell Dirac-Fock
 state $|\Phi\rangle$. Whereas $S_{\text{v}}$ excites at least one electron from the valence orbital, giving rise to valence and core-valence excited configurations \cite{Lindgren1987}.  The operator $S_{\text{v}}$ can yield the valence and core-valence excited configurations with respect to the open-shell Dirac–Fock state $|\Phi_\text{v}\rangle$ \cite{Dutta2016}. Here we generate single- and double-excited correlated configurations from Eq. (1).  The amplitudes of these excitations are solved from the energy eigenvalue equations of the closed-shell and open-shell systems, which are $He^T|\Phi\rangle =Ee^T|\Phi\rangle$
 and $H_\text{v}e^T|\Phi_\text{v}\rangle= E_\text{v}e^T|\Phi_\text{v}\rangle$, respectively \cite{Dixit2007a}. In the present method, these amplitudes are solved following the Jacobi iteration scheme, which is considered all-ordered. The initial guesses of the single- and double-excitation amplitudes are made consistent with the first order of the perturbation theory \cite{Lindgren1985}. In the present version of RCC theory, we also consider some important triple excitations and hence the abbreviation is used RCCSD(T).
	
 The matrix elements of an arbitrary operator can be written as
	\begin{eqnarray}
	O_{ki}& = & \frac{\langle{\Psi_k|\hat{O}|\Psi_i}\rangle}{\sqrt{\langle{\Psi_k|\Psi_k}\rangle{\langle{\Psi_i|\Psi_i}\rangle}}}\nonumber\\
	& = &\frac{\langle{\Phi_k|\{1+S^{\dagger}_k\}e^{T^{\dagger}}{\hat{O}}e^T\{1+S_i\}|\Phi_i}\rangle}{\sqrt {{\langle{\Phi_k|\{1+S^{\dagger}_k\}e^{T^{\dagger}}e^T\{1+S_k\}|\Phi_k}\rangle\langle\Phi_i|\{1+S^{\dagger}_i\}e^{T^{\dagger}}e^T\{1+S_i\}|\Phi_i}\rangle}}.
	\end{eqnarray}
	
	The detailed derivations and explanations of the matrix elements associated with electric dipole ($E_1$), electric quadrupole ($E_2$), 
 and magnetic dipole ($M_1$) transitions can be found in the literature\cite{Savukov2003}. Emission transition probabilities (s$^{-1}$) for the $E_1$, $E_2$, and $M_1$ from $|\Psi_k\rangle$ to $|\Psi_i\rangle$ state are  \cite{Shore1968}
	\begin{equation}
\hspace{1cm}	A^{E_1}_{k\rightarrow i}=\frac{2.0261\times10^{-6}}{\lambda^3(2J_k+1)}S^{E_1},
	\end{equation}
	\begin{equation}
\hspace{1cm}	A^{E_2}_{k\rightarrow i}=\frac{1.12\times10^{-22}}{\lambda^5(2J_k+1)}S^{E_2},
	\end{equation}
	\begin{equation}
and \hspace{1cm}	A^{M_1}_{k\rightarrow i}=\frac{2.6971\times10^{-11}}{\lambda^3(2J_k+1)}S^{M_1}.
	\end{equation}
	Where, $\lambda$ is in cm and $S$ is the
	square of the transition matrix elements of $O$ (corresponding transition operator) in atomic unit of $e^2a_0^2$($e$ is the charge of an electron and $a_0$ is the Bohr radius). The oscillator strength for the $E1$ transition is related to the corresponding transition probability (s$^{-1}$) with following equation \cite{Kelleher2008}
	\begin{equation}
	f^{osci}_{k\rightarrow i}= 1.4992\times10^{-16}A_{k\rightarrow i}\frac{g_k}{g_i}\lambda^2,
	\end{equation}
	where $g_k$ and $g_i$ are the degeneracies of the final and initial states, respectively. 
	The lifetime of the $k$-th state is calculated by considering all transition probabilities to the lower energy states ($i$-th)  and is given by 
\begin{equation}
\tau_k = \frac{1}{\sum_i {A_{k\rightarrow i}}}.
\end{equation}
	In order to incorporate the plasma screening effect on the atomic spectroscopic properties, the Dirac-Coulomb potential takes the form as
 \begin{equation}
	H^{D}_{\mathrm{eff}}=H+V^{D}_{\mathrm{eff}}({\bf r}_{i})
 \end{equation}
	Here  $V^{D}_{\mathrm{eff}}({\bf r}_{i})$ is the effective potential of the nucleus on the $i$-th  electron due to the presence of the plasma environment. The Debye-H\"{u}ckle potential is considered to examine the effect of screening of nuclear coulomb potential due to the presence of ions and free electrons in plasma \cite{Ichimaru1982, Akhiezer1982}. In the case of a weekly interacting plasma medium, the effective potential experienced by the $i$-th electron is given as 
   \begin{equation}
	V^{D}_{\mathrm{eff}}({r}_{i})=\frac{Ze^{-\mu{ r}_{i} }}{{ r_{i}}}. 
 \end{equation}
	Where Z is the nuclear charge and $\mu$ is the Debye screening parameter, which is related to the ion density $n_{ion}$ and plasma temperature T through the following relation
 \begin{equation}
	\mu=\left[{\frac{4\pi(1+Z)n_{ion}}{K_{B}T}}\right]^2,
 \end{equation}
	where $k_B$ is the Boltzmann constant. Therefore, a given value of $\mu$ represents a range of plasma conditions with different ion densities and temperatures.
	
		\section{Results and discussions}
	The single-particle Dirac-Fock (DF) wavefunctions are the building blocks of the RCC calculations yielding the many-electron correlation energies and  correlated wavefunctions. We calculate the bound Dirac-Hartree Fock orbitals as accurately as possible using a sophisticated numerical  approach, GRASP92 \cite{Parpia2006}. Further, we apply the basis-set expansion  technique  \cite{Clementi1990} in the self-consistent field approach to obtain Gaussian-type DF orbital (GTO) used in the RCC calculations.  The radial part of each basis function has two parameters, $\alpha_0$, and $\beta$, as exponents \cite{Huzinaga1993} to be optimized. The parameters are required to optimize due to the finite size of the basis set. The exponent parameters are optimized compared to the DF bound orbitals obtained from GRASP92, discussed in detail in our old papers \cite{Dutta2013, Roy2015}. In the basis optimization method, we consider 33, 30, 28, 25, 21,  and 20 basis functions for $s$, $p$, $d$, $f$, $g$,  and $h$ symmetries, respectively. This basis set is considered for all the ions. But the choice of the active orbitals in the RCC calculation relies on the convergence of the correlation contribution to the closed-shell energy with the increasing number of the orbitals \cite{Roy2014, Roy2015}. Therefore, the active orbitals for the converged correlation contribution to the closed-shell energy are found to be distinct for different ions investigated in this work.
	
	 In this article, we calculate the ionization potential of Ar$^{7+}$, Kr$^{7+}$, Xe$^{7+}$,  and Rn$^{7+}$ using the RCC method and compare them in Table 1 with results published in National Institute of Standards and Technology (NIST) \cite{Kramida2019} wherever available. NIST estimations are considered to have the best accuracy.  We find that our calculated ground state energies of Ar$^{7+}$, Kr$^{7+}$, and Xe$^{7+}$  are in excellent agreement with NIST results, and deviations are estimated to be $-0.01\%$, $0.45\%$, and $-0.03\%$, respectively. Table I presents the ionisation potential of the low-lying excited states of these ions with average deviations around $-0.08\%$, $0.42\%$, and $0.30\%$, respectively. In these cases, the maximum difference is $-0.23\%$ occurred for $5p_{3/2}$ state of Ar$^{7+}$, $0.60\%$ for 
	$5g_{7/2, 9/2}$ of Kr$^{7+}$, and $1.2\% $ for $6d_{3/2,5/2}$ of Xe$^{7+}$. 
	 
	 Our calculated energies agree well with estimations by Fischer et al. \cite{Fischer2006}, who computed energy levels of Ar$^{7+}$ using the core polarization effect on the Dirac Hartree-Fock (CP-DHF) theory. Cheng and Kim \cite{Cheng1978} tabulated the energy levels of Kr$^{7+}$ from the relativistic Hartree-Fock (RHF) calculations. As expected, our RCC calculated results are found to be better in agreement with the NIST values. For Rn$^{7+}$, we have not found any experimental measurement in the literature nor NIST compiled values. Only one theoretical calculation based on the CP-DHF method by \cite{Migdalek2018} is available with an average deviation of IP is $0.66\%$ from our calculations. 
	
	 The percentage of electron correlation correction, i.e.,  $\frac{(\text{RCC-DF})\times 100\%}{\text{DF}}$ in IP of the ground state is monotonically increases from Ar$^{7+}$ to Rn$^{7+}$ with the values $0.39\%$, $0.52\%$, $1.81\%$,  and $1.87\%$, respectively. \\

\begin{figure}
 \begin{subfigure}{0.5\textwidth}
		\centering
		\includegraphics[width=0.8\linewidth]{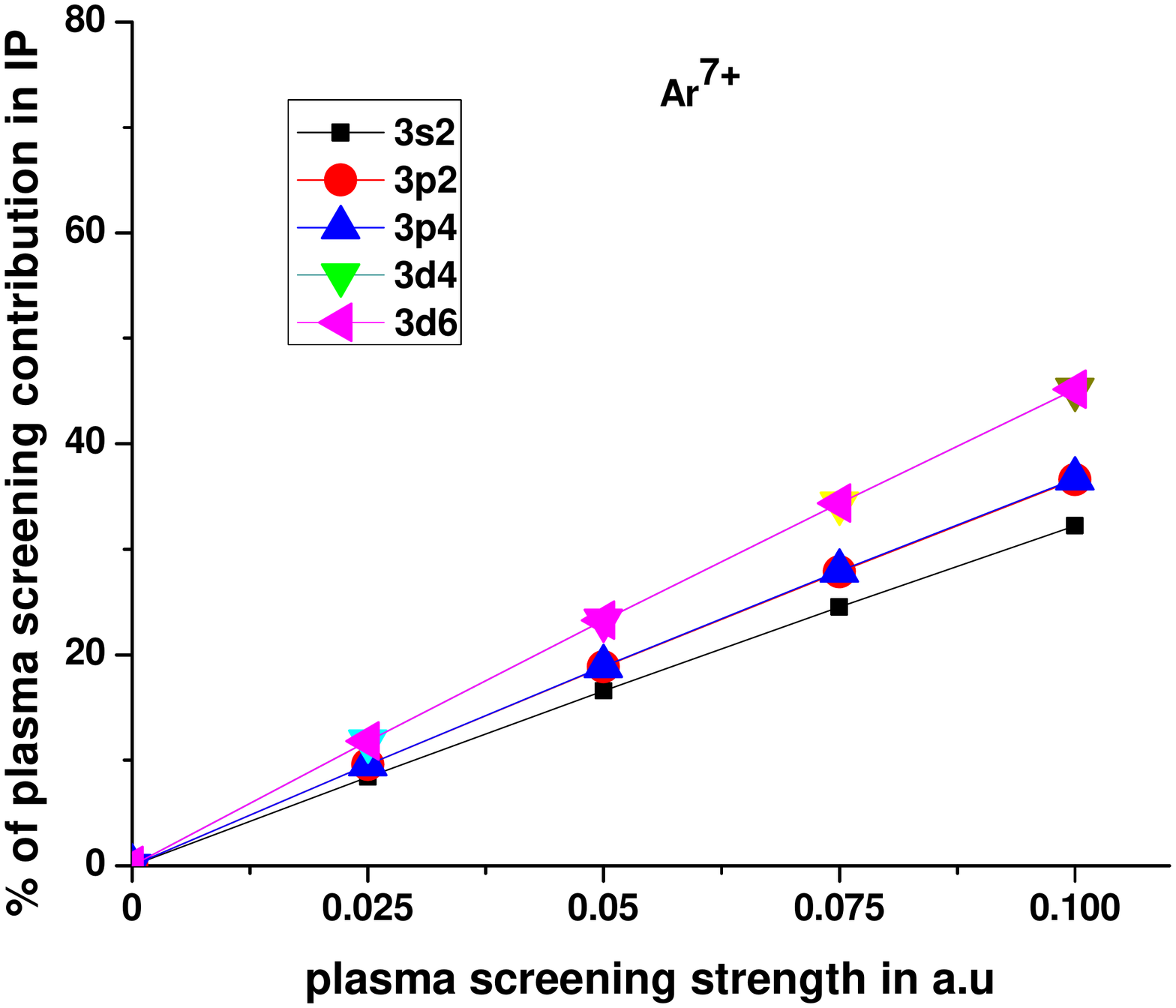}

	\end{subfigure}
	\begin{subfigure}{0.5\textwidth}
		\centering
		\includegraphics[width=0.8\linewidth]{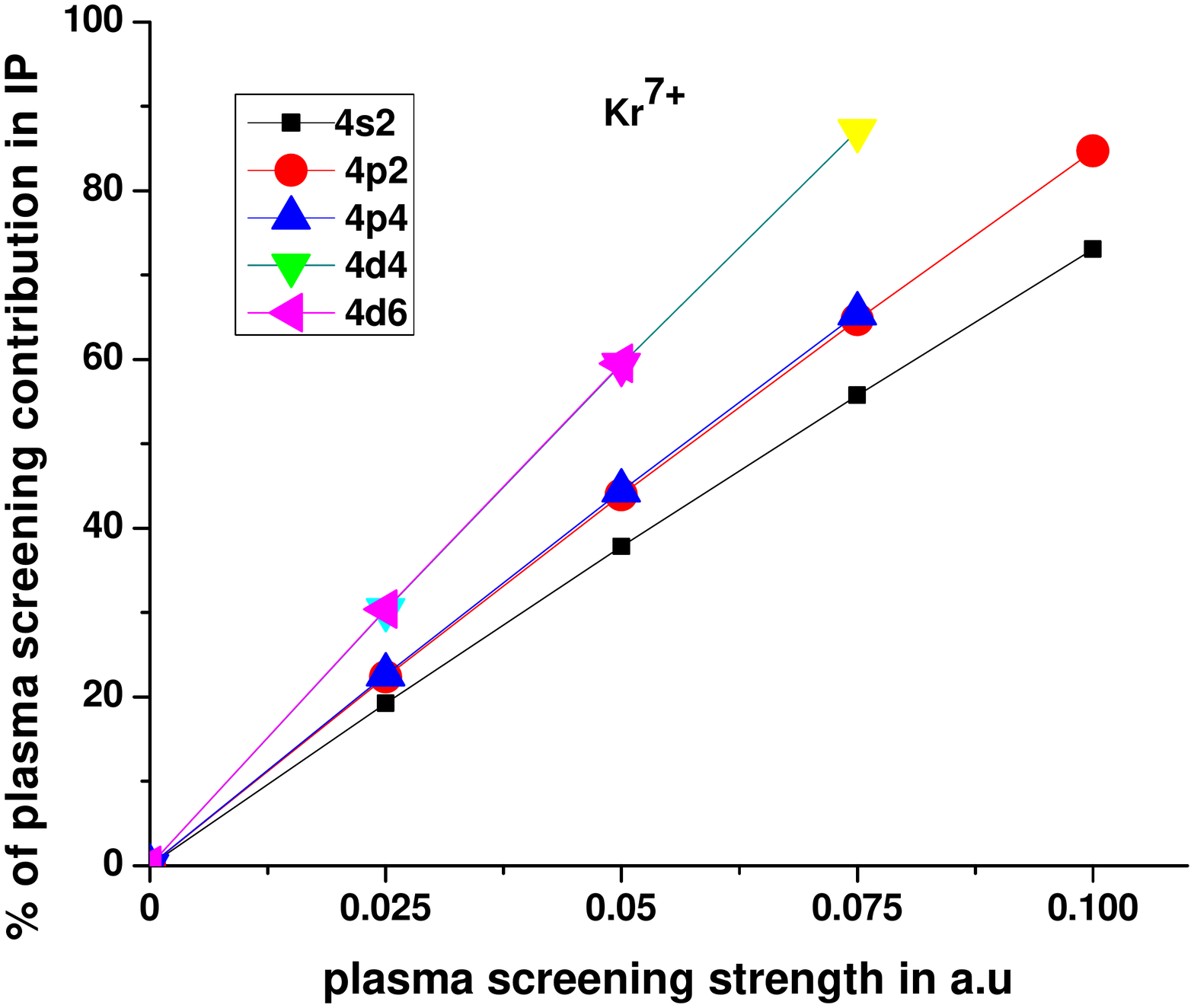}

	\end{subfigure}
	\begin{subfigure}{0.5\textwidth}
		\centering
		\includegraphics[width=0.8\linewidth]{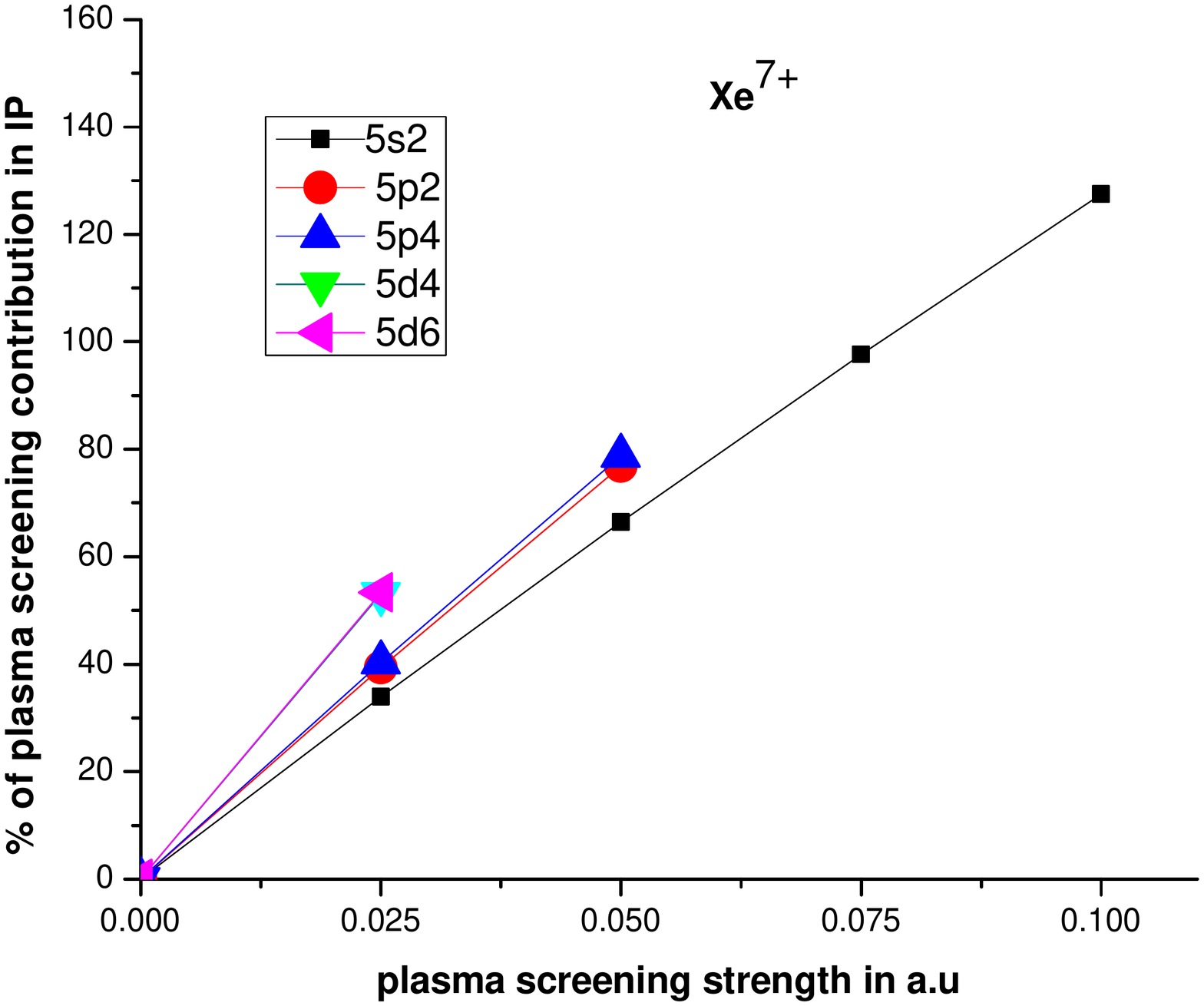}

	\end{subfigure}
	\begin{subfigure}{0.5\textwidth}
		\centering
		\includegraphics[width=0.8\linewidth]{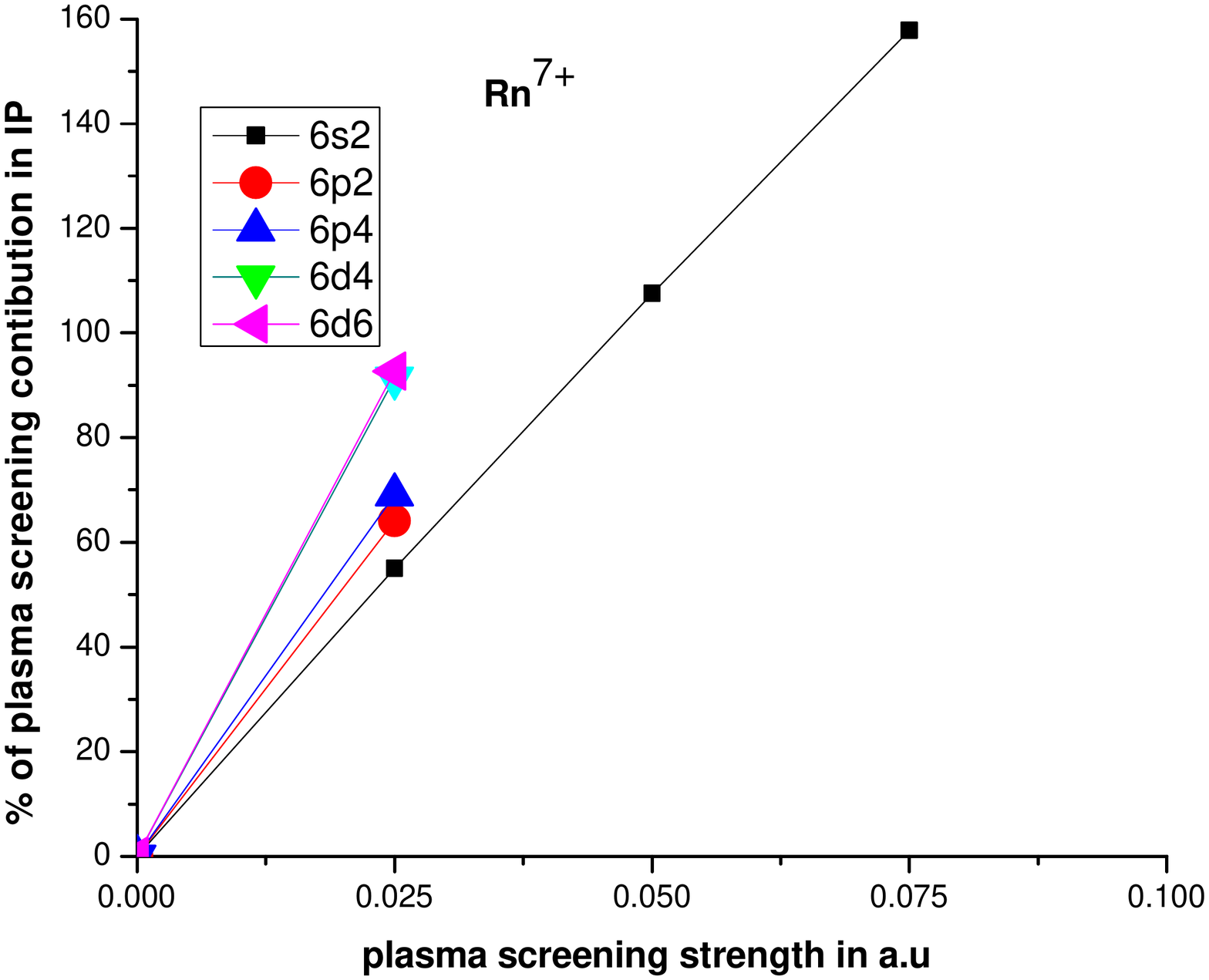}
	
	\end{subfigure}
	\label{fig:sfig1}
\caption{The plasma contribution in IP of low-lying septuple ionized atoms: \% variation of IP with plasma screening strength. Energy levels are indicated as $nL(2J+1)$. Results are calculated from $\frac{\text{IP}_{\mu=0}-\text{IP}_{\mu >  0}}{\text{IP}_{\mu=0}}\times 100$.}	
\end{figure}

Now we investigate the impact of the plasma screening potential on the energy levels of the considered ions. Table I shows that IP monotonically decreases with the increase of the $\mu$ value. The bold values for each ion in the table represent the limiting case beyond which the system becomes unbound. Figure 1 presents the plasma screening contribution in IP for a few low-lying states, such as ground state $S_{1/2}$, excited $P_{1/2,3/2}$, and $D_{3/2,5/2}$ states of Ar$^{7+}$, Kr$^{7+}$, Xe$^{7+}$, and Ra$^{7+}$ ions. The panels of the figure show the plasma screening contribution increases from the ground to higher excited states, as the latter states are less bound by the Coulomb attraction.  For Xe$^{7+}$ and Rn$^{7+}$, we could plot the effect up to a certain value of $\mu$ as most of the states become continuum states beyond that. We observe that the plasma screening effect is practically strong for fine structure levels for  Ar$^{7+}$ and Kr$^{7+}$ ions and weak for Xe$^{7+}$, and Ra$^{7+}$ ions.

  \begin{figure}
	\begin{subfigure}{0.5\textwidth}
		\centering
		\includegraphics[width=1\linewidth]{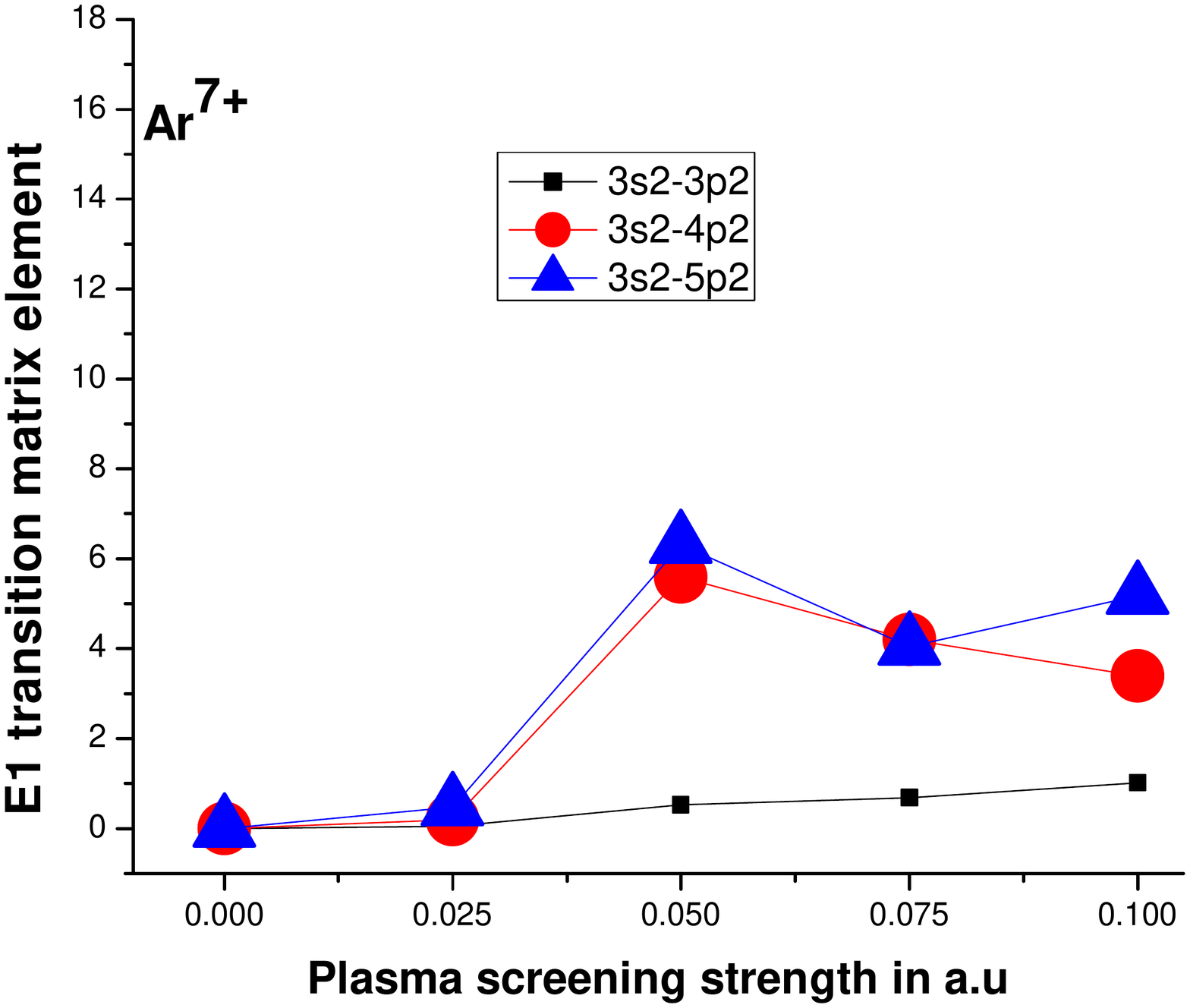}

	\end{subfigure}
	\begin{subfigure}{0.5\textwidth}
		\centering
		\includegraphics[width=1\linewidth]{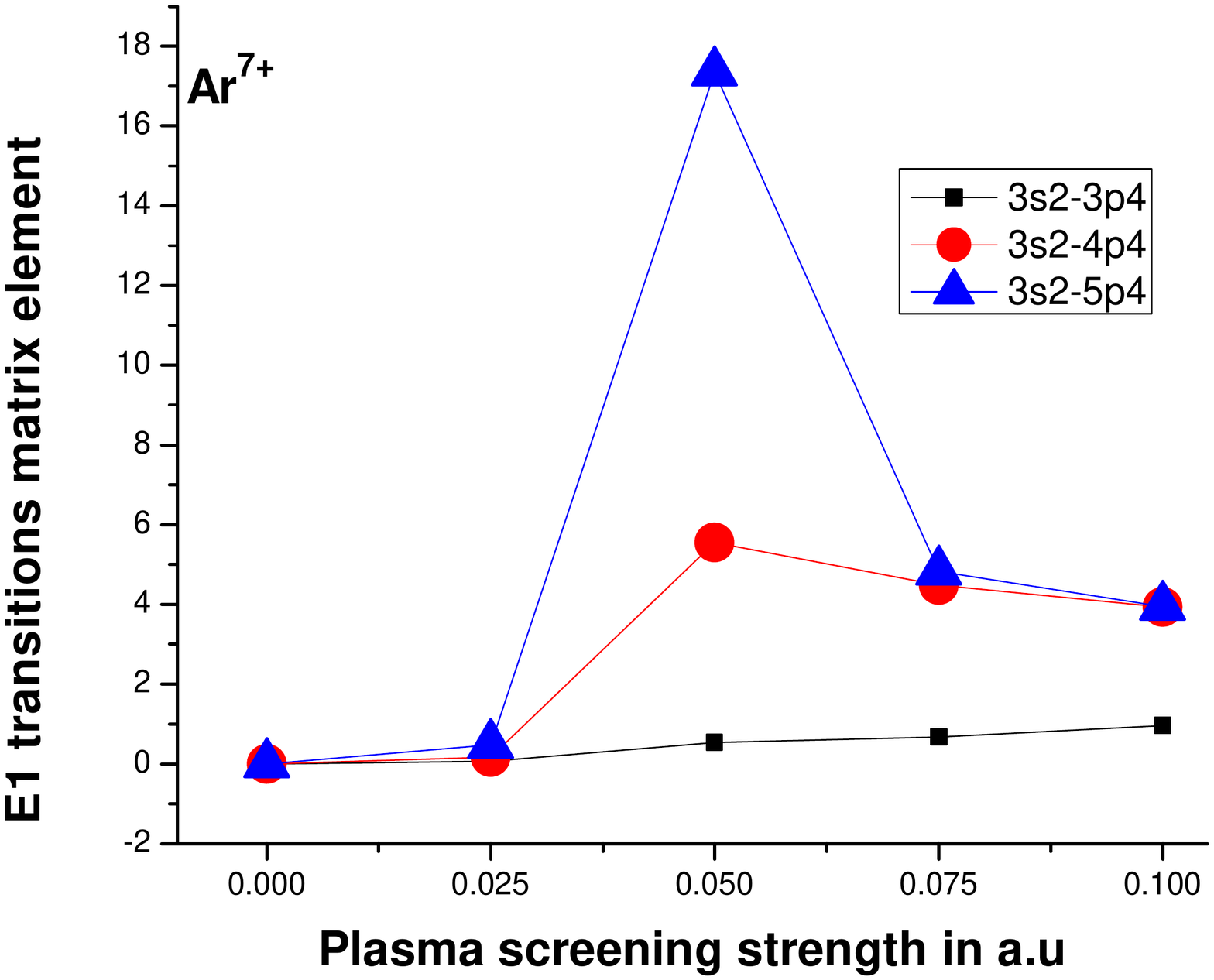}

	\end{subfigure}	
        \label{fig:sfig1e1}	
		\caption{The contribution of plasma screening in E1 matrix elements of Ar$^{7+}$ of transitions $^2S_{1/2}\rightarrow$ $^3P_{1/2,3/2}$. The figures display \% variation of the E1 matrix elements  vs. plasma screening strength.}
\end{figure}
	
	We present the electric dipole matrix elements for the ions in the plasma medium in Table 2. The table also displays  our computed DF values of the matrix elements to reveal the correlation contributions. The separate presentation of the DF and RCC values over the span of the plasma screening parameter, $\mu$, in the table is an intentional move. Here we want to highlight that the plasma screening impacts the DF and the RCC correlation parts differently. In the case of Ar$^{7+}$, the average changes in matrix element due to the increasing values of plasma screening parameter  is less than 1$\%$. However, Figure. 2 shows significant changes for $3s_{1/2}\rightarrow 4p_{1/2, 3/2}$ and  $3s_{1/2}\rightarrow 5p_{1/2, 3/2}$ for finite values of $\mu$, especially for $\mu=0.05$.  This is true for any $n\ ^2S_{1/2}$ $\rightarrow$ $n^\prime\ ^2P_{1/2,3/2}$ of this ionic series. However, apart from such a few transitions,  the  plasma effects lie between 1 $\%$ to 2 $\%$ for most of the other transitions in the series.  Table 2 shows that the average correlations (for the non-plasma environment) in the transition amplitudes for Ar$^{7+}$, Kr$^{7+}$,  Xe$^{7+}$, and Rn$^{7+}$ are $0.3858 \%$, $4.0831\%$, $7.6074\%$, $11.6379\%$ apart from $4d_{3/2}$$\rightarrow$ $5f_{5/2}$ transition where the correlation is $67.11 \%$.

	For observational astronomy and laboratory spectroscopy, we present a tabulation of a list of our computed oscillator strengths ($f_{\text{RCC}}$) of $E1$ transitions along with their previously reported theoretical and experimental values in Table 3. Most of the transitions fall in the far and mid-UV regions of the electromagnetic spectrum. $f_{\text{RCC}}$ is calculated using the RCC transition amplitudes in length gauge \cite{Dutta2013a} form presented in Table 2 and the  NIST \cite{Kramida2019} wavelengths, wherever available (in the case of Rn$^{7+}$, our computed RCC  wavelengths are used). The ratios between the length and velocity gauge amplitudes of our calculated $E1$-transitions are also displayed in table to show the accuracy of our RCC wavefunctions, which is close to unity for all the cases confirming the accuracy of our correlated atomic wavefunctions. However, we find that the ratio is almost two for $^2F\rightarrow ^2D$ transitions of  Xe$^{7+}$ and Rn$^{7+}$. Point to note that this disagreement is also available in the ratio at the  DF level,  where we also employed the numerically accurate GRASP92 Code \cite{Parpia2006}.  One of the reasons for this outcome is due to the strong correlation effect from d- and f-states, so as in similar alkali systems \cite{Jyoti2022}. In addition, the consistency of accuracy of our calculations can be drawn from the approximate consistency of the ratios, 3:2:1  among the transition matrix estimations of $^2P_{3/2} \rightarrow$ $^2D_{5/2}$: $^2P_{1/2} \rightarrow$ $^2D_{3/2}$: $^2P_{3/2} \rightarrow$ $^2D_{3/2}$ \cite{Cowan1981}.

 The $E1$ Oscillator strengths for Ar$^{7+}$ are well studied in literature \cite{Reistad1986, Verner1996, Feret1999, Lagmago1998, Siegel1998, Guet1990, Theodosiou1988, Kim1978, Fischer2006, Gruzdev1979}, and they are in good  agreement with our estimations based on correlation exhaustive RCC method.  Table 3 shows that the same is true for Kr$^{7+}$.  For Ar$^{7+}$,  our calculations for $f_{\text{RCC}}$ are almost as accurate as those found from other sophisticated theoretical approaches, such as the relativistic many-body perturbation theory \cite{Guet1990},  and for the most latest theoretical results employing the multiconfigurational Dirac-Hartree-Fock approximation \cite{Fischer2006}. To the best of our knowledge, in the case of Kr$^{7+}$, we could not find any correlation-exhaustive many-body  result of $E1$ transition. There have been experiments, mostly using beam-foil experiments, on the $E1$ transition from the ground state to the first excited states of  Ar$^{7+}$ \cite{Reistad1986}, and Kr$^{7+}$ \cite{Irwin1976, Pinnington1979,Knystautas1977, Livingston1980, Druetta1976}. Our estimations are well within the uncertainty limit of the latest experiments. We also see that some of the old calculations either underestimate or overestimate the oscillator strength values due to non-appropriate considerations of correlations and relativistic effects.

	Over the last two decades, a few of the low-lying $E1$ transitions of Xe$^{7+}$ are estimated using core-polarization or model potential as an effective means of correlation calculations, apart from third order perturbation calculations \cite{Safronova2003}. It is known that our RCC method is an all-order extension of many-body perturbation theory \cite{Lindgren1985}. Further, it includes most of the correlation features, including core correlation, pair correlation, and higher order correlation effects \cite{Dutta2016} for a given level of excitation. For Rn$^{7+}$, we find only one theoretical endeavour \cite{Migdalek2018} using model potential. The presence of $d$- and $f$-orbitals for Xe$^{7+}$ and Rn$^{7+}$ ions in the core makes these two ions highly correlated. Because of the large atomic number and highly stripped  configurations, we expect a strong relativistic effect in their spectroscopy. Therefore, it is necessary to do relativistic {\it ab initio} correlation exhaustive calculations for them and our computations exactly mitigate that requirement. In Table 3, we also present the effect of the plasma atmosphere on the oscillator strengths for the observational and laboratory spectroscopy. The oscillator strengths for $\mu>0$ are calculated using the $E1$ matrix elements presented in Table 2 and the corresponding transition wavelengths computed from RCC theory. The table exhibits the significant effects of plasma screening parameters on the oscillator strengths.

	Table 4 and Table 5 present transition probabilities for the relatively strong forbidden transitions governed by the electric quadruple ($E2$) and the magnetic dipole ($M1$) moments. Similar to oscillator strength in Table 3, here we use NIST wavelengths for the transition probability wherever available. For Rn$^{7+}$, we use the RCC calculated transition wavelengths. We do not find any estimation of the forbidden transitions in the literature of this ionic series which fall either in the ultraviolet or in the near infra-red regions of the electromagnetic spectrum.   Transitions falling in the ultraviolet region are significant in astronomical observation and plasma research \cite{Saloman2004, Morita2010, Fahy2007, Morgan1995}. While the infra-red transitions have applications in astronomy using space-based telescopes  (\cite{Kessler1996}). Moreover, infrared spectroscopy provides major information about cool astronomical regions in space, like, interstellar medium \cite{Feuchtgruber1997} and  planetary nebulae \cite{Liu2001}. It is found that $5p_{1/2}$ $\rightarrow$ $4f_{5/2}$ of Kr$^{7+}$ and $5p_{1/2}$ $\rightarrow$ $5p_{3/2}$ of Xe$^{7+}$ emit orange and green lights, respectively, which can be used in laser spectroscopy.

    It is found from Table 5 that the $M1$ transition probability is stronger among fine-structure levels than the $E2$ transition. Table 4  reveals that the maximum $M1$ transition probability, $A_{RCC}^{M1}$, occurs for the transition  $3p_{1/2}$ $\rightarrow$ $3p_{3/2}$ of Ar$^{7+}$, $4p_{1/2}$ $\rightarrow$ $4p_{3/2}$  of Kr$^{7+}$, $4f_{7/2}$ $\rightarrow$ $5f_{7/2}$ of Xe$^{7+}$ and $6p_{1/2}$ $\rightarrow$ $6p_{3/2}$ of Rn$^{7+}$, and they have  values 0.17949, 8.4000, 158.33 and 1137.7, respectively. Moreover, our estimations of $M1$ transition probability for the $4f_{5/2}$ $\rightarrow$ $4f_{7/2}$ transition of Xe$^{7+}$ has excellent agreement with  the calculations using the multi-configuration Dirac-Hartree-Fock method \cite{Grumer2014, Ding2012}. Table 6 presents lifetime of the low-lying states of this series. We compare our results with other experimental and theoretical estimations wherever available and find good agreement with the recent endeavors. We provide lifetimes of many excited states calculated the first time in literature to our knowledge.\\
	
	The comparisons of our computed results with the other estimations obtained from correlation exhaustive {\it ab initio} theoretical computations or precise experiments are the one of the measures of accuracy of our calculations. Further, the differences between the calculated matrix elements in the length and velocity gauge forms are the characteristic of the preciseness of our calculations. A recent literature \cite{Jyoti2022} also claims that the difference in length gauge and velocity gauge is a measure of accuracy.   Another factor of accuracy in {\it ab initio} calculations  arises from the DF wavefunctions used for correlation calculations.  Also, we should consider the uncertainty that arises from the other correlation terms (which we did not consider in this article) and the quantum electrodynamics effect, which is at most $ 2 \%$ in total. Taking all these into account, the maximum calculated uncertainties for Ar$^{7+}$, Kr$^{7+}$, Xe$^{7+}$, and  Rn$^{7+}$ are about $5.6\%$, $5.37\%$, $5\%$ and $5.01\%$, respectively. \\

	To understand the critical effect of the plasma atmosphere on the ionization potential of the ions, we highlight the IPD values in bold fonts in Table 1 for different values of screening length, $\mu$.  These IPD values reflect critical electron or plasma density at a particular temperature for the ionic system when a few of the bound ionic states are elevated to continuum states. 
	
	Table 1 also reveals  that the fine structure splittings (FSS) are suppressed as the screening strength increases from $\mu$=0 to 1.0.  For example, the energy differences between $4p_{3/2} $ and $4p_{1/2}$ of Kr$^{7+}$ are evaluated as  9774 a.u, 9732 a.u, 9608 a.u., and 9407 a.u. for  $\mu$ = 0, 0.025, 0.05, 0.075 a.u., respectively. This phenomenon is consistent with earlier calculations for sodium D line \cite{Basu2014}, and hydrogen-like atoms  \cite{Xie2014}.  The suppression of the transition rate among the fine-structure levels are mainly arising from the energy quench.
	
	From Figure 3, we pictorially estimate the critical values of plasma screening strength ($\mu_{c}$) where the ionisation potential becomes zero for particular atomic state. We also tabulate these value in Table 7. The critical screening strength is essential in photo-ionisation cross section, which increases with increasing $\mu$ till $\mu=\mu_{c}$. This increment is obvious due to the decrease of bound state energy leading to the increase of radial expansion of bound state wavefunction \cite{Xie2014}. This phase shift of bound state to continuum state is induced by the plasma atmosphere, and ionization threshold decreases with the Debye screening length ($\mu^{-1}$). In terms of photo-ionization cross-section \cite{Sahoo2009}, plasma decreases the threshold  cross-section, and the discrete bound wavefunctions become diffused. Therefore, critical screening strength plays a important role in atomic structure.  However, we have not found any spectroscopic data in literature for these ions in plasma medium to compare with our results.
 
	\section{conclusion}
	The continuous progress in astrophysical and astronomical observations demands accurate theoretical transition data in a realistic environment. In many cases, the experiment is difficult to extract the data used to estimate the abundance of the ions in the stellar chemical composition. Here, the highly correlated relativistic coupled-cluster theory is applied to precisely determine the excitation energies of a few low-lying states of astrophysically relevant such as  Ar$^{7+}$,  Kr$^{7+}$ and Xe$^{7+}$ and Rn$^{7+}$. Furthermore, we calculate various properties of allowed and forbidden transitions, such as transition probabilities, oscillator strengths and lifetimes, and compare them with previously reported data in the literature. We found an overall good agreement between our results with the other theoretical and experimental results. Moreover, the concurrence between the length and velocity gauge allowed transition amplitudes signifies the exact calculations of our correlated wavefunctions. We found that most of the transitions shown here fall in the ultraviolet region of the electromagnetic spectrum, useful for  astrophysical plasma research and telescope-based astronomy. A few transitions, such as $4d_{3/2}-4f_{5/2}$, $4d_{5/2}-4f_{5/2}$ and  $4d_{5/2}-4f_{7/2}$ of Ar$^{7+}$, $5p_{1/2}-4f_{5/2}$ of   Kr$^{7+}$  and $5p_{1/2}-5p_{3/2}$ of Xe$^{7+}$ emit the visible light which can have application in laser spectroscopy. Our presented transition line parameters of Rn$^{7+}$ may help the astronomer identify the ion's unknown lines.  To the best of our knowledge, some of the oscillator strengths of allowed transitions and most of the transition rates of the forbidden transitions are reported here  for the first time in the literature. 

    The main focus of this paper is to evaluate the above spectroscopic properties under a realistic astronomical atmosphere. We showed the variation of our results for different values of Debye screening lengths and ionisation potential depression values for each atomic state useful for atomic structure characterization. 

    \clearpage
			\LTcapwidth=\linewidth
		\begin{longtable}{c c c c c c c c c  }

		\caption{Comparison of our RCC ionization potential (in $cm^ {-1}$)  with NIST data and our estimations of plasma screening effect on them. Estimations for $5g$ states of  Rn$^{7+}$ were not available in the literature (a) \cite{Migdalek2018}. Plasma screening strength ($\mu$) is in a.u. unit. Energy levels are indicated as $nL(2J+1)$. The bold values indicate that beyond which the system becomes unbound. } \\
			\hline\hline
			
			state&NIST &$\mu$=0&$\mu$=0.025 &$\mu$=0.05   & $\mu$=0.075 & $\mu$=0.1\\
			\hline
			
			Ar$^{7+}$  & & & & && \\
			3s2 &1157056&1157201 &1059866 &965330&873513&784345   \\
			3p2 &1016961 &1016995 &919704 &825299&733697&644828  \\
			3p4 &1014248 &1014184 &916898& 822505&730924&642084 \\
			3d4&824447 &824210 &726923& 632529&540956&452148 \\
			3d6&824302 &824027 &726741& 632349&540779&451974 \\
			4s2&581098 &581069&485070& 394397&308793&228054 \\
			4p2&528815 &528528 &432658& 342356&257356&177460 \\
			4p4&527813 &527428 &431565&341287&256323&176473  \\
			4d4&459524 &459475 &363709 &273714&189240&110134  \\
			4d6&459435 &459386 &363620 & 273627&189156&110055 \\
			4f6&440204 &440190 &344137 & 253339&167618&86891 \\
			4f8&440181 &440159 &344107 & 253309&167590&86865 \\
			5s2&349750 &349752 &255504 & 169770&91961&21742 \\
			5p2&324795 &323193 &229236 & 144332&67862&\textbf{557} \\
			5p4&324307 &322543 &228611 & 143769&67376 &  \\
			5d4&291782&291699 &197867 &113352 & 37637&  \\
			5d6&291778&291652 &197818 &113302 & 37591&  \\
			5f6&281727&281736 &187657 &102487 & 25846& \\
			5f8&281707&281720 &187642 &102473 & 25833& \\
			5g8&281051 &281015 &186389 & 99684&20685 \\
			5g10&281037 &281015 &186380 & 99676&20678& \\
			Kr$^{7+}$  & & & && \\
			4s2&1014665 &1010099 &815902 &628237 &446944&271892 \\
			4p2&870969 &867027 &673080 & 486149&306059&132689 \\
			4p4&861189 &857253 &663348 & 476541&296652&123557 \\
			4d4&640619 &636965 &443698 & 258762&81891& \textbf{86678}\\
			4d6&639288 &635514 &442257 & 257352&80623& \\
			5s2&524578 &523198 &331842 & 152413&\textbf{15482}& \\
			5p2&467984&465448 &274592&96629&&\\
			5p4&464221 &461406 &270630 & 92891&& \\
			4f6&451900 &450180 &257831 & 75655&& \\
			4f8&451934 &450186 &257837 & 75660&& \\
			5d4&373589 &371922 &182111 & 7230&& \\
			5d6&373048 &371273 &181476 & 6635&& \\
			6s2&322147 &321635 &133957 &\textbf{34847}&& \\
			5f6&289666&288637&100173&&&\\
            5f8&289661&288634 &100170&&&\\
			5g8&281574 &281387 &92121 & && \\
			5g10&281572 &281379 &92114 & && \\
			Xe$^{7+}$  & & & && \\	
			5s2&854769 &854995 &564858 &286731 &20272& \\
			5p2&738302 &737059 &447391 &170643 &\textbf{93515}& \\
			5p4&719717 &718263 &428759 &152487 && \\
			4f6&589608 &588730 &297599 &16588 & & \\
			4f8&589058 &588088 &296970 &16085 && \\
			5d4&544881 &543506 &255203  &\textbf{17518} && \\
			5d6&541953 &540549 &252295  & && \\
			6s2&459272 &455364 &170155  & && \\
			6p2&411391 &406318 &121973 & & &\\
			6p4&403996 &398801 &114717 & & & & \\
			5f6&357190 &356376 &70683 & && \\
			5f8&356751 &355922 &70245 & && \\
			6d4&327344 &323450 &41320 & && \\
			6d6&325975 &322176 &40102 & && \\
			7s2&289473 &276902 &\textbf{451} & && \\
			5g8&284501 &283609 & & && \\
			5g10&284501 &283617 & & && \\
			Rn$^{7+}$  &(a) & & && \\	
			6s2&834624 &839362 &377923 &\textbf{63552} && \\
			6p2&712821 &718015 &257667 & && \\
			6p4&661071 &665627 &205616 & && \\
			5f6&536092 &531079 &69545 & && \\
			5f8&534720 &529291 &68106 & && \\
			6d4& 498636&501235 &43424 & && \\
			6d6&491240 &493699 &36116 & && \\
            7s2&446847 &445577 &\textbf{8041} & && \\
            7p2&397883& 398535&& && \\
            7p4&377435 &377332 & & && \\
			5g8& &287997 & & && \\
			5g10& &288252 & & && \\
			\hline
		\end{longtable}	
  \clearpage

	\LTcapwidth=\linewidth
	\begin{longtable}{ c c  c c c c c c c  c c}
		\caption{ Our DF and  RCC matrix element (a.u.), in length gauge,  of electric dipole ($E1$) transitions in plasma medium. Energy levels are indicated as $nL(2J+1)$.  } \\
		
		\hline\hline 
		&	\multicolumn{2}{c}{$\mu=0$} &\multicolumn{2}{c}{$\mu=0.025$}
		&\multicolumn{2}{c}{$\mu=0.05$}&\multicolumn{2}{c}{$\mu=0.075$}&\multicolumn{2}{c}{$\mu=0.1$}\\
		
	Ar$^{7+}$   &  &  &  &  &  &  &  &  &  &  \\ \hline
		transitions & DF & RCC & DF & RCC & DF & RCC & DF & RCC & DF & RCC \\ \hline
		3s2$\rightarrow$3p2	 & 0.9617 & 0.9341 & 0.9623 & 0.9346 & 0.9638 & 0.9390 & 0.9663 & 0.9405 & 0.9697 & 0.9436 \\ 
		3s2$\rightarrow$3p4 & 1.3619 & 1.3228 & 1.3626 & 1.3236 & 1.3648 & 1.3299 & 1.3683 & 1.3317 & 1.3732 & 1.3355 \\ 
		3s2$\rightarrow$4p2 & 0.2002 & 0.2093 & 0.1999 & 0.2089 & 0.1989 & 0.2210 & 0.1973 & 0.2181 & 0.1950 & 0.2164 \\ 
		3s2$\rightarrow$4p4 &0.2770 & 0.2898&   0.2765&  0.2893 & 0.2751 & 0.3059&  0.2728&  0.3028&   0.2698&  0.3012 \\ 
		3s2$\rightarrow$5p2 & 0.1011 & 0.1062 & 0.1006 & 0.1057 & 0.0992 & 0.0995 & 0.0972 & 0.1105 & 0.0944 & 0.1007 \\ 
		3s2$\rightarrow$5p4 &0.1397 & 0.1470 &  0.1390&  0.1463&  0.1372&  0.1215&   0.1345&  0.1541& 0.1307 &0.1528  \\
		4s2$\rightarrow$4p2 & 1.8958 & 1.8813 & 1.8996 & 1.8851 & 1.9105 & 1.9059 & 1.9284 & 1.8987 & 1.9537 & 1.9012 \\ 
		4s2$\rightarrow$4p4 & 2.6827 & 2.6622 & 2.6879 & 2.6675 & 2.7033 & 2.7006 & 2.7286 & 2.6857 & 2.7644 & 2.6844 \\ 
		3p2$\rightarrow$4s2 &0.3604 & 0.3646 & 0.3608 & 0.3649 & 0.3619 & 0.3419 & 0.3637 & 0.3437 & 0.3661 & 0.3432 \\ 
		3p4$\rightarrow$4s2 & 0.5172 & 0.5231 & 0.5178 & 0.5236 & 0.5193 & 0.4898 & 0.5219 & 0.4943 & 0.5253 & 0.4952 \\ 
		3p2$\rightarrow$5s2 & 0.1237 & 0.1269 &0.1240  & 0.1271 & 0.1246 & 0.1275 & 0.1252& 0.1144 & 0.1253 & 0.1274      \\ 
        3p4$\rightarrow$5s2 & 0.1771 & 0.1816 &0.1775& 0.1819 & 0.1783 & 0.2021 &0.1792 & 0.1653 & 0.1793 & 0.1659 \\
		3p2$\rightarrow$3d4 & 1.3534 & 1.3174 & 1.3547 & 1.3186 & 1.3583 & 1.3207 & 1.3642 & 1.3218 & 1.3725 & 1.3305 \\ 
		3p4$\rightarrow$3d4 & 0.6060 & 0.5899 & 0.6066 & 0.5905 & 0.6082 & 0.5916 & 0.6109 & 0.5918 & 0.6146 & 0.5956 \\ 
		3p4$\rightarrow$3d6 & 1.8184 & 1.7700 & 1.8201 & 1.7720 & 1.8250 & 1.7736 & 1.8330 & 1.7750 & 1.8442 & 1.7795 \\ 
		3p2$\rightarrow$4d4 & 0.3857 & 0.3957 & 0.3842 & 0.3942 & 0.3800 & 0.3925 & 0.3730 & 0.3926 & 0.3629 & 0.3551 \\ 
		3p4$\rightarrow$4d4 & 0.1756 & 0.1801 & 0.1750 & 0.1794 & 0.1731 & 0.1770 & 0.1699 & 0.1791 & 0.1654 & 0.1645 \\ 
		3p4$\rightarrow$4d6 & 0.5262 & 0.5394 & 0.5242 & 0.5375 & 0.5185 & 0.5321 & 0.5091 & 0.5372 & 0.4956 & 0.5357 \\ 
		4p2$\rightarrow$4d4 & 2.7891 & 2.7687 & 2.7966 & 2.7761 & 2.8181 & 2.7737 & 2.8536 & 2.7737 & 2.9038 & 2.8032 \\ 
		4p4$\rightarrow$4d4&  1.2496 & 1.2405 & 1.2529 & 1.2438 & 1.2626 & 1.2470 & 1.2785 & 1.2420 & 1.3010 & 1.2498 \\ 
		4p4$\rightarrow$4d6 & 3.7490 & 3.7217 & 3.7590 & 3.7316 & 3.7880 & 3.7880 & 3.8357 & 3.7255 & 3.9033 & 3.7279 \\ 
		3d4$\rightarrow$4p2 & 0.5411 & 0.5434 & 0.5426 &0.5449  &0.5470  & 0.5616 & 0.5542 & 0.5407 & 0.5644 & 0.5701 \\ 
		3d4$\rightarrow$4p4 & 0.2395 & 0.2406 &0.2402 & 0.2412 & 0.2487 & 0.2450 & 0.2454 & 0.2395 &0.2499  & 0.2530 \\ 
		3d6$\rightarrow$4p4 & 0.7193 & 0.7225 & 0.7213 &0.7244 & 0.7271 & 0.7193 & 0.7368 & 0.7201 & 0.7504  &0.7184 \\ 
		3d4$\rightarrow$4f6 & 1.7707 & 1.7378 & 1.7705 & 1.7377 & 1.7702 & 1.7384 & 1.7691 & 1.7401 & 1.7663 & 1.7778 \\ 
		3d6$\rightarrow$4f6 & 0.4734 & 0.4671 & 0.4734 & 0.4646 & 0.4733 & 0.4649 & 0.4730 & 0.4653 & 0.4722 & 0.4658 \\ 
		3d6$\rightarrow$4f8 & 2.1172 & 2.078 & 2.1171 & 2.0779 & 2.1166 & 2.0793 & 2.1154 & 2.0768 & 2.1121 & 2.0799 \\ 
		4d4$\rightarrow$4f6 & 3.1546 & 3.1495 & 3.1686 & 3.1634 & 3.2100 & 3.1508 & 3.2796 & 3.1504 & 3.3822 & 3.1414 \\ 
		4d6$\rightarrow$4f6 & 0.8430 & 0.8416 & 0.8468 & 0.8454 & 0.8578 & 0.8420 & 0.8764 & 0.8419 & 0.9087 & 0.8423 \\ 
		4d6$\rightarrow$4f8 & 3.7706 & 3.7644 & 3.7874 & 3.7812 & 3.7652 & 3.7644 & 3.9201 & 3.7778 & 4.0427 & 3.7790 \\ 
		
		\hline
	
		Kr$^{7+}$  &  &  &  &  &  &  &  &  &  &  \\ \hline
		~  &DF& RCC & DF & RCC & DF & RCC & DF & RCC & DF & RCC \\ \hline
		4s2$\rightarrow$4p2 & 1.1348 & 1.0794 & 1.1363 & 1.1017 & 1.1405 & 1.1223 & 1.1475 & 1.1088 & 1.1575 & 1.1245 \\ 
		4s2$\rightarrow$4p4 & 1.6095 & 1.5314 & 1.6115 & 1.5632 & 1.6175 & 1.5900 & 1.6275 & 1.5724 & 1.6416 & 1.6085 \\ 
		4s2$\rightarrow$5p2 & 0.1488 & 0.1655 & 0.1479 & 0.1584 & 0.1451 & 0.1537 &  &  &  & \\ 
		4s2$\rightarrow$5p4 & 0.1812 & 0.2057 & 0.1799 & 0.1944 & 0.1761 & 0.1839 &  &  &  &  \\ 
		4p2$\rightarrow$4d4 & 1.7539 & 1.6812 & 1.7569 & 1.7412 & 1.7657 & 1.7506 &  1.7804 & 1.7422   &  &  \\ 
		4p4$\rightarrow$4d4 & 0.7940 & 0.7616 & 0.7954 & 0.7734 & 0.7995 & 0.7836 &  0.8064 & 0.7771  &  &  \\ 
		4p4$\rightarrow$4d6 & 2.3818 & 2.2847 & 2.3860 & 2.3218 & 2.3984 & 2.3511 & 2.4190 & 2.3301   &  &  \\
		4p2$\rightarrow$5s2 & 0.4884 & 0.4934 & 0.4894 & 0.4892 & 0.4925 & 0.4838 &  &  &  &  \\ 
		4p4$\rightarrow$5s2 & 0.7316 & 0.7379 & 0.7331 & 0.7325 & 0.7377 & 0.7272 &  &  &  &  \\ 
		4p2$\rightarrow$6s2 & 0.1655 & 0.1686 & 0.1654 & 0.1664 & 0.1646 & 0.1624 &  &  &  &  \\ 
		4p4$\rightarrow$6s2 & 0.2447 & 0.2486 & 0.2445 & 0.2460 & 0.2427 & 0.2424 &  &  &  &  \\ 
		4d4$\rightarrow$4f6 & 2.8243 & 2.7466 & 2.8340 & 2.7759 & 2.8623 & 2.7985 &  &  &  &  \\ 
        4d6$\rightarrow$4f6 &0.7565  &0.7358  &0.7591  & 0.7436 &0.7668  & 0.7499 &  & &  &  \\
        
        4d4$\rightarrow$5p2 &1.1740  & 1.1630 &1.1809  & 1.1776 &1.2015  &1.1742  &  & &  &  \\
        4d4$\rightarrow$5p4 &  0.5090 & 0.5124 &0.5120  & 0.5107 &0.5212  &0.5168  &  &  &  &  \\
        4d6$\rightarrow$5p4 & 1.5415 &1.5522  &1.5507 &1.5464  & 1.5786  & 1.5435 &  &  &  &  \\
        
		4d6$\rightarrow$4f8 &3.3836 & 3.2916 & 3.3953 & 3.3257 & 3.4297 & 3.3529 &  &  &  &  \\ 
        4d4$\rightarrow$5f6 &0.0596 & 0.0427  & 0.0582&   0.0445 &   &  &  &  &  \\
        
		4d6$\rightarrow$5f6 & 0.0133 & 0.0091 & 0.0175 & 0.0089 & &  &  &  & \\
	    4d6$\rightarrow$5f8 & 0.0592 & 0.0418 & 0.0779 & 0.0394 & &  &  &  & \\	 
		4f6$\rightarrow$5g8 & 3.8576 & 3.7993 & 3.8688 & 3.8112 &  &  &  &  &  &  \\ 
		4f8$\rightarrow$5g8 & 0.7422 & 0.7311 & 0.7444 & 0.7340 &  &  &  &  &  &  \\ 
		4f8$\rightarrow$5g10& 4.3914 & 4.3256 & 4.4042 & 4.3380 &  &  &  &  &  &  \\ 
		4f6$\rightarrow$5d4 & 2.0421 & 2.0387 & 2.0728 & 2.0318 & 2.1683 & 2.0704 &  &  &  & \\

        4f6$\rightarrow$5d6 & 0.5430 &  0.5420 & 0.5512 &0.5401  & 0.5766 &0.5498  &  &  &  & \\
  
		4f8$\rightarrow$5d6 & 2.4274 & 2.4232 & 2.4641 & 2.4137 & 2.5780 & 2.4567 &  &  &  & \\ 
		5p2$\rightarrow$5d4 & 3.1683 & 3.1275 & 3.1867 & 3.1373 & 3.2401 & 3.1494 &  &  &  &  \\ 
		5p4$\rightarrow$5d4 & 1.4407 & 1.4226 & 1.4489 & 1.4271 & 1.4728 & 1.4325 &  &  &  &  \\ 
		5p4$\rightarrow$5d6 & 4.3172 & 4.2629 & 4.3417 & 4.2781 & 4.4131 & 4.2912 &  &  &  & \\ 
        5p2$\rightarrow$6s2 &1.0220 & 1.0239  &1.0277  &1.0215 &  &  &  &  & \\
        5p4$\rightarrow$6s2 &1.5278 & 1.5310  &1.5347  & 1.5276&  &  &  &  & \\
		\hline
	
		&  &  &  &  &  &  &  &  &  &  \\ \hline
		Xe$^{7+}$ & DF  & RCC & DF & RCC & DF & RCC & DF & RCC & DF & RCC \\ \hline
		5s2$\rightarrow$5p2 & 1.3758 & 1.1736 & 1.3791 & 1.1769 & 1.3889 & 1.1866 &  &  & &\\ 
		5s2$\rightarrow$5p4 & 1.9516 & 1.6705 & 1.9562 & 1.6752 & 1.9700 & 1.6890 &  &  &  &\\ 
		5p2$\rightarrow$5d4 & 2.1135 & 1.8629 & 2.1197 & 1.8697 &  &  &  &  &  &  \\ 
		5p4$\rightarrow$5d4 & 0.9768 & 0.8646 & 0.9799 & 0.8679 &  &  &  &  &  &  \\ 
		5p4$\rightarrow$5d6 & 2.9252 & 2.5911 & 2.9344 & 2.6011 &  &  &  &  &  &  \\ 
		5p2$\rightarrow$6s2 & 0.5866 & 0.6026 & 0.5893 & 0.6046 &  &  &  &  &  &  \\ 
		5p4$\rightarrow$6s2& 0.9449 & 0.9582 & 0.9494 & 0.9619 &  &  &  &  &  &  \\ 
		
		4f6$\rightarrow$5d4 & 1.7952 & 1.5958 & 1.8177 & 1.6155 &  &  &  &  &  &  \\ 
		4f6$\rightarrow$5d6 & 0.4761 & 0.4237 & 0.4822 & 0.4291 &  &  &  &  &  &  \\ 
		4f8$\rightarrow$5d6 & 2.1392 & 1.9075 & 2.1663 & 1.9313 &  &  &  &  &  &  \\ 
		4f6$\rightarrow$5g8 & 1.8079 & 1.5869 & 1.7885 & 1.5731 &  &  &  &  &  &  \\ 
		4f8$\rightarrow$5g8 & 0.3495 & 0.3073 & 0.3457 & 0.3046 &  & &  &  &  &  \\ 
		4f8$\rightarrow$5g10 & 2.0684 & 1.819 & 2.0462 & 1.8032 &  &  &  &  &  &  \\ 
		5d4$\rightarrow$5f6 & 2.9732 & 2.8328 & 2.9687 & 2.8347 &  &  &  &  &  &  \\ 
		5d6$\rightarrow$5f6& 0.8050 & 0.7670 & 0.8040 & 0.7676 &  &  &  &  &  &  \\ 
		5d6$\rightarrow$5f8 & 3.5922 & 3.4233 & 3.5876 & 3.4252 &  &  &  &  &  &  \\ 
		
		6s2$\rightarrow$6p2 &2.7436&2.5326&2.5640&2.5931&  &  &  &  &  &  \\ 
		6s2$\rightarrow$6p4 &3.3346& 3.5052& 3.6133& 3.6569 & &  &  &  &  &  \\ 
		6p2$\rightarrow$6d4 &3.7253&3.5930&3.7520& 3.6089 & &  &  &  &  &  \\ 
		6p4$\rightarrow$6d4&1.7277&1.6725&1.7399& 1.6762 & &  &  &  &  &  \\ 
		6p4$\rightarrow$6d6&5.1635&4.9966&5.1994& 5.0147& &  &  &  &  &  \\ 
		5f6$\rightarrow$5g8& 5.5926 & 5.3626 &  &  &  &  &  &  &  &  \\ 
        5f6$\rightarrow$5g8&1.0772  &1.0330  &  &  &  &  &  &  &  &  \\
		5f8$\rightarrow$5g10 & 6.3725 & 6.1114 &  &  &  &  &  & &  &  \\ 
		\hline
	
		&  &  &  &  &  &  &  &  &  &  \\ \hline
		Rn$^{7+}$ & DF & RCC & DF & RCC & DF & RCC & DF & RCC & DF & RCC \\ \hline
		6s2$\rightarrow$6p2 & 1.4159 & 1.1344 & 1.4210 & 1.1414 &  &  &  &  &  &  \\ 
		6s2$\rightarrow$6p4 & 1.9902 & 1.6147 & 1.9969 & 1.6264 &  &  &  &  &  &  \\ 
		6p2$\rightarrow$7s2 & 0.6184 & 0.6284 & 0.6189 & 0.6678 &  &  &  &  &  &  \\ 
		6p4$\rightarrow$7s2 & 1.2512 & 1.2162 & 1.2657 & 1.2024 &  &  &  &  &  &  \\ 
		6p2$\rightarrow$6d4 & 2.0887 & 1.7562 & 2.0978 & 1.7629 &  &  &  &  &  &  \\ 
		6p4$\rightarrow$6d4 & 1.0493 & 0.8953 & 1.0556 & 0.9008 &  &  &  &  &  &  \\ 
		6p4$\rightarrow$6d6 &3.1200 & 2.6596 & 3.1380 & 2.6691 &  &  &  &  &  &  \\ 
		5f6$\rightarrow$6d4 & 2.3362 & 2.0900 & 2.3794 & 2.0911 &  &  &  &  &  &  \\ 
		5f6$\rightarrow$6d6& 0.6118 & 0.5500 & 0.6236 & 0.5478 &  &  &  &  &  &  \\ 
		5f8$\rightarrow$6d6& 2.7709 & 2.4962 & 2.8234 & 2.4931 &  &  &  &  &  &  \\ 
		5f6$\rightarrow$5g8& 2.7042 & 2.3829 &  &  &  &  &  &  &  &  \\ 
		5f8$\rightarrow$5g8& 0.5263 & 0.4650 &  &  &  &  &  &  &  &  \\ 
        5f8$\rightarrow$5g10 & 3.1174 & 2.7455 &  &  &  &  &  &  &  &  \\
		7s2$\rightarrow$7p2 &2.5750  &2.4195 &  &  &  &  &  &  &  &  \\
        7s2$\rightarrow$7p4 & 3.5648 &3.3636  &  &  &  &  &  &  &  &  \\
		6d4$\rightarrow$7p2 &2.0098  &1.9415  &  &  &  &  &  &  &  &  \\
        6d4$\rightarrow$7p4 &0.7525  &0.7381  &  &  &  &  &  &  &  &  \\
        6d6$\rightarrow$7p4 &2.4112  & 2.3331 &  &  &  &  &  &  &  &  \\

		\hline	
	\end{longtable}	
	
\clearpage
	
\LTcapwidth=\linewidth
\begin{longtable}{c c  l c c c c c}
	\caption{ Our RCC oscillator strengths of electric dipole transitions. We compare our results with other estimations available in recent literature (experimental endeavours are highlighted with "exp" subscript). Our results ("RCC") are obtained using the RCC calculations, except NIST wavelengths are used for $\mu=0$ wherever available. Transition states are designated with the outermost orbital followed by $(2J+1)$ of the state. Values at the parenthesis in the second column are ratios between length- and velocity-gauged dipole matrix elements. } \\
	
	\hline\hline 	
	 &   \multicolumn{2}{c}{$\mu$=0 } &$\mu$=0.025&$\mu$=0.050&$\mu$=0.075&$\mu$=0.1 \\
	\cline{2-3} 
	Transition  &$ RCC $ & Other&&&& \\  
	\hline
	\endfirsthead
	
	\multicolumn{7}{c}%
	{{\bfseries \tablename\ \thetable{} --Continued from previous page}} \\
	\hline
	 & \multicolumn{2}{c}{$\mu$=0 }&$\mu$=0.025&$\mu$=0.05&$\mu$=0.075&$\mu$=0.1  \\
	\cline{2-3} 
	Transition  & RCC & Other &&&& \\
	\hline
	\endhead
	
	\hline \multicolumn{3}{c}{Continued on next page} \\
	\endfoot
	\hline
	\endlastfoot
	
	Ar$^{7+}$ \\
	3s2 $\rightarrow$ 3p2 &0.1857(0.99)&  0.183(4)$^{a_1}_{exp}$, $0.188^{b_1}$, &0.1859&0.1875&0.1878&0.1887 \\
	&  &$0.193^{c_1}$, $0.186^{d_1}$, $0.1864^{e_1} $,   &&&& \\
	&  & $0.185^{f_1,i_1}$, $0.187^{g_1}$, $0.196^{h_1}$ &&&&\\
	3s2 $\rightarrow$ 3p4 & 0.3795(0.99) &0.398(10)$^{a_1}_{exp}$, 0.385$^{b_1}$, 0.394$^{c_1}$, &0.3804&0.3837&0.3840&0.3854\\
	&   & 0.381$^{d_1, g_1}$, $0.3811^{e_1}$, $0.379^{f_1}$,  &&&& \\
	& &    $0.401^{h_1}$, $0.378^{i_1}$ &&&& \\
	3s2 $\rightarrow$ 4p2&0.0418(1.00) &$0.0414^{b_1}$, $0.0401^{c_1}$, $0.0376^{d_1}$&0.0416&0.0462&0.0445&0.0432  \\
  &  &$0.0415^{e_1}$,  $0.0385^{h_1}$, $0.0432^{i_1}$ \\
	3s2 $\rightarrow$ 4p4&0.0803(1.00) &$0.0829^{b_1}$, $0.0766^{c_1}$, $0.0751^{d_1}$, &0.0793&0.0877&0.0859&0.0838\\
   &  &$0.0798^{e_1}$,  $0.0739^{h_1}$, $0.0836^{i_1}$&&&&\\
	3s2 $\rightarrow$ 5p2 &0.0143(0.98) &$0.0146^{b_1, e_1}$&0.0141&0.0123&0.0149&0.0121 \\
	3s2 $\rightarrow$ 5p4 &0.0271(1.00) &$0.0292^{b_1}$, $0.0284^{e_1}$&0.0270&0.0184&0.0291&0.0278 \\
	4s2 $\rightarrow$ 4p2 &0.2810(1.01) & $0.2821^{e_1}$, $0.2824^{i_1}$&0.2829&0.2871&0.2816&0.2777 \\
	4s2 $\rightarrow$ 4p4 &0.5736(1.01) &$0.5755^{e_1} $,   $0.5759^{i_1}$&0.5782&0.5883&0.5748&0.5645 \\
	3p2 $\rightarrow$ 4s2 &0.0880 (1.00)&$0.0876^{e_1}$, $0.0884^{i_1}$&0.0879&0.0765&0.0763&0.0760\\
	3p4 $\rightarrow$ 4s2 &0.0900(1.01) &$0.0896^{e_1}$, $0.08947^{i_1}$&0.0899&0.0780&0.0783&0.0783\\
	3p2 $\rightarrow$ 5s2 &0.0163(1.01) &$0.0161^{e_1}$ &0.0163&0.0162&0.0128&0.0154 \\
    3p4 $\rightarrow$ 5s2 &0.0166(1.01) &$0.0164^{e_1}$ &0.0166&0.0202&0.0133&0.0130 \\
	3p2 $\rightarrow$ 3d4 &0.5074(0.96) &$0.532^{c_1}$, $0.5097^{e_1}$, $0.5074^{g_1}$&0.5091&0.5107&0.5115&0.5180 \\
	&  &  $0.508^{i_1}$, $0.47^{j_1}$&&&& \\
	3p4 $\rightarrow$ 3d4 &0.0502(0.96) &$0.0527^{c_1}$,  $0.0504^{e_1}$, $0.0501^{g_1}$&0.0503&0.0505&0.0505&0.0512   \\
	&  &  $0.0502^{i_1}$, $0.046^{j_1}$&&&& \\
	3p4 $\rightarrow$ 3d6 &0.4556(0.96) &$0.475^{c_1}$, $0.4539^{e_1}$, $0.4517^{g_1}$&0.4534&0.4542&0.4549&0.4571    \\
	&  &  $0.452^{i_1}$, $0.42^{j_1}$&&&& \\
	3p2 $\rightarrow$ 4d4 &0.1326(1.05) & $0.1310^{e_1} $, $0.1344^{i_1}$&0.1312&0.1291&0.1275&0.1024  \\
	3p4 $\rightarrow$ 4d4 &0.0137(1.03) &$0.0135^{e_1} $, $0.0136^{i_1}$&0.0135&0.0131&0.0132&0.0109\\
	3p4 $\rightarrow$ 4d6 &0.1226(1.03) &$0.1212^{e_1} $, $0.1228^{i_1}$&0.1214&0.1180&0.1187&0.1159\\
	4p2 $\rightarrow$ 4d4 &0.8067(0.95) &$0.8085^{i_1}$&0.8070&0.8020&0.7930&0.8035 \\
	4p4 $\rightarrow$ 4d4 &0.0798(0.98) &$0.0796^{i_1}$&0.0797&0.0798&0.0786&0.0787 \\
	4p4 $\rightarrow$ 4d6 &0.7192(0.98) &$0.7177^{i_1}$&0.7185&0.7215&0.7079&0.7009 \\
	3d4 $\rightarrow$ 4p2 &0.0663(0.98) &$0.0657^{e_1}$,  $0.0663^{i_1}$&0.0663&0.0695&0.0630&0.0678 \\
	3d4 $\rightarrow$ 4p4 &0.0130(1.00) &$0.0129^{e_1}$,   $0.0131^{i_1}$&0.0130&0.0133&0.0124&0.0134 \\
	3d6 $\rightarrow$ 4p4 &0.0784(0.97) &$0.0778^{e_1}$,  $0.0786^{i_1}$&0.0784&0.0762&0.0747&0.0720 \\
	3d4 $\rightarrow$ 4f6 &0.8812(1.00) &$0.8776^{i_1}$&0.8777&0.8702&0.8584&0.8762 \\
	3d6 $\rightarrow$ 4f6 &0.0424(1.00) &$0.0418^{i_1}$&0.0418&0.0415&0.0409&0.0401 \\
	3d6 $\rightarrow$ 4f8 & 0.8397(1.00)&$0.8360^{i_1}$&0.8364&0.8297&0.8149&0.7796 \\

	 Kr$^{7+}$\\
	4s2 $\longrightarrow $ 4p2   &0.2543(1.00) & 0.25(1)$^{a_2,f_2}_{exp}$, 0.24(2)$^{b_2}_{exp}$, 0.246$^{c_2}$,   &0.2633 &0.2718&0.2705&0.2673 \\
	&  &  0.2781$^{d_2}$, 0.278$^{e_2}$, $0.28^{g_2}$,  &&&& \\
	&  &0.2578$^{h_2}$, 0.220$^{i_2}$, 0.2448$^{j_2}$&&&& \\
	4s2 $\longrightarrow $ 4p4 &0.5466(0.97) &0.53$(2)^{a_2}_{exp}$, 0.47$(4)^{b_2}_{exp}$,  0.526$^{c_2}$,  &0.5661&0.5825&0.5833&0.5829 \\
	& &0.5965$^{d_2}$, 0.60$^{e_2}$, 0.59$(9)^{f_2}_{exp}$,   &&&& \\
	& &$0.59^{g_2}$,  $0.554^{h_2}$, $0.473^{i_2}$&&&& \\
    & &0.5265$^{j_2} $  &&&&\\
	4s2 $\longrightarrow $ 5p2 &0.0227(1.01)  &$0.0176^{d_2}$   &      0.0206&     0.0191&    && \\	
	4s2 $\longrightarrow $ 5p4&0.0354(1.00)& $0.0265^{d_2}$         &0.0313&       0.0275&   & \\ 
	4p2 $\rightarrow$ 5s2 &0.1281(1.02)  &$0.1212^{d_2}$&0.1261&0.1187&&\\
	4p4 $\rightarrow$ 5s2 &0.1392(1.02)  &$0.1321^{d_2}$&0.1351&0.1302&&\\
	4p2 $\rightarrow$ 4d4 &0.9888(1.00)  &$1.057^{d_2}$&1.0562&1.0583&1.1068&\\
	4p4 $\rightarrow$ 4d4 &0.0972(1.00)  &$0.1038^{d_2}$&0.0998&0.1015&0.1045&\\
	4p4 $\rightarrow$ 4d6 &0.8796(1.00)  &$0.9395^{d_2}$&0.9051&0.9200&0.9480&\\
	4p2 $\rightarrow$ 6s2 &0.0237(1.06)  &$0.0208^{d_2}$&0.0227&0.0181&&\\ 
	4p4 $\rightarrow$ 6s2 &0.0253(1.07)  &$0.0220^{d_2}$&0.0243&0.0197&&\\
	4d4 $\rightarrow$ 4f6 &1.0811(0.98)  &$1.126^{d_2}$&1.0876&1.0890&& \\
	4d6 $\rightarrow$ 4f6 &0.0514(0.98)  &$0.0535^{d_2}$&0.0508&0.0509&& \\
	4d6 $\rightarrow$ 4f8 &1.0277(0.98)  &$1.0700^{d_2}$&1.0326&1.0340&& \\
	4d4 $\rightarrow$ 5p2 &0.1828(1.00)  &$0.1784^{d_2}$&0.1781&0.1697&& \\
	4d4 $\rightarrow$ 5p4 &0.0352(1.03)  &$0.0344^{d_2}$&0.0345&0.0329&&\\
	4d6 $\rightarrow$ 5p4 &0.2135(1.00)  &$0.2088^{d_2}$&0.2078&0.1984&& \\
	4d4 $\rightarrow$ 5f6 &0.0001(1.09)  &$0.0001^{d_2}$&0.0005&&& \\
	4f6 $\rightarrow$ 5d4 &0.1647(0.98)  &$0.1638^{d_2}$&0.1582&0.1485&& \\
	4f6 $\rightarrow$ 5d6 &0.0117(0.98)  &$0.0117^{d_2}$&0.0113&0.0106&& \\
	4f8 $\rightarrow$ 5d6 &0.1759(0.98)  &$0.1750^{d_2}$&0.1689&0.1582&& \\
	4f6 $\rightarrow$ 5g8 &1.2447(1.00)  &$1.261^{d_2}$&1.2185&&& \\
	4f8 $\rightarrow$ 5g8 &0.0346(1.00)  &$0.0350^{d_2}$&0.0339&&& \\
	4f8 $\rightarrow$ 5g10 &1.2103(1.00) &$1.226^{d_2}$&1.1841&&& \\
	5p2 $\rightarrow$ 5d4 &1.4023(1.00)  &$1.401^{d_2}$&1.3825&1.3467&&\\
	5p4 $\rightarrow$ 5d4 & 0.1393(1.01) &$0.1383^{d_2}$&0.1369&0.1335&& \\
	5p4 $\rightarrow$ 5d6 &1.2582(1.01)  &$1.249^{d_2}$&1.2391&1.2062&&\\
    4d6 $\rightarrow$ 5f6 &0.00002(0.28)  &$0.00003^{d_2}$&0.00001&&& \\
    4d6 $\rightarrow$ 5f8 &0.0003(0.29)  &$0.0007^{d_2}$&0.0003&&& \\
    5p2 $\rightarrow$ 6s2 &0.2322(1.07)  &$0.2023^{d_2}$&0.2229&&&\\
    5p4 $\rightarrow$ 6s2 &0.2529(1.08)  &$0.2179^{d_2}$&0.2422&&&\\
	Xe$^{7+}$ \\
	5s2 $\rightarrow $ 5p2 & 0.2436(1.01)&$0.294^{a_3}$, $0.234^{b_3}$, $0.242^{c_3}$&0.2471&0.2482&&\\  
	&  &$0.253^{d_3}$, $0.237^{e_3}$, $0.237^{f_3}$&&&&\\
    &  & $0.232^{g_3}$, $0.223^{h_3}$, $0.232^{i_3}$&&&& \\
	5s2 $\rightarrow $ 5p4&0.5724(1.01) &$0.697^{a_3}$, $0.550^{b_3}$, $0.569^{c_3}$&0.5801&0.5816&&\\ 
	&   &$0.596^{d_3}$, $0.560^{e_3}$, $0.563^{f_3}$, &&&&  \\
	&   &   $0.543^{g_3}$, $0.522^{h_3}$, $0.537^{i_3}$ &&&& \\ 
	5p2 $\rightarrow$5d4 &1.0195(1.02) &$1.189^{a_3}$, $0.977^{b_3}$, $1.020^{c_3}$&1.0204&&&  \\
	& & $1.025^{d_3}$, $1.003^{e_3}$, $1.000^{f_3}$&&&& \\
	& &  $1.057^{i}$ &&&& \\
	5p4 $\rightarrow$5d4 &0.0992(1.02) &$0.095^{b_3}$, $0.089^{c_3}$, $0.099^{d_3}$&0.0993&&& \\
	&  &$0.097^{e_3}$, $0.097^{f_3}$, $0.095^{i_3}$&&&& \\
	5p4 $\rightarrow$5d6 &0.9064(1.02) &$0.523^{a_3}$, $0.868^{b_3}$, $0.904^{c_3}$&0.9066&&&  \\
    &  &$0.907^{d_3}$, $0.889^{e_3}$, $0.886^{f_3}$&&&&\\
	&  &$0.875^{i_3}$&&&& \\
	5p2 $\rightarrow$ 6s2 &0.1539(1.05) &$0.160^{c_3}$, $0.156^{d_3}$, $0.155^{e_3}$&0.1539&&& \\
	&  &$0.153^{f_3}$, $0.199^{i_3}$&&&& \\ 
	
	5p4 $\rightarrow$ 6s2 &0.1816(1.05) &$0.188^{c_3}$, $0.186^{d_3}$, $0.184^{e_3}$&0.1817&&&\\
	&  &  $0.182^{f_3}$, $0.186^{i_3}$&&&& \\
	4f6 $\rightarrow $5d4 &0.0577(2.35) &$0.130^{a_3}$, $0.058^{b_3}$, $0.060^{i_3}$&0.0560&&& \\
	4f6 $\rightarrow $5d6 &0.0043(2.12) &$0.0044^{b_3}$&0.0042&&& \\
	4f8 $\rightarrow $5d6 &0.0651(2.19) &$0.075^{a_3}$, $0.065^{b_3}$, $0.068^{i_3}$&0.0633&&& \\
	4f6 $\rightarrow$ 5g8  &0.3890(1.02) &$0.3646^{b_3}$, $0.354^{i_3}$&&&& \\	
	4f8 $\rightarrow$ 5g8 &0.0109(1.02) &$0.0102^{b_3}$&&&&  \\
	4f8 $\rightarrow$ 5g10 &0.3826(1.02) &$0.3595^{b_3}$, $0.343^{i_3}$&&&& \\
	5d4 $\rightarrow $5f6 &1.1437(1.04) &$1.099^{i_3}$&1.1259&&& \\
	5d6 $\rightarrow $5f6  &0.0550(1.04) &$0.052^{i_3}$&0.0813&&& \\
	5d6 $\rightarrow $5f8 &1.0981(1.04) &$1.032^{i_3}$&1.0812&&& \\
	5f6 $\rightarrow$ 5g8 &1.0583(0.99) &$1.071^{i_3}$&&&& \\
	5f8 $\rightarrow$ 5g8 &0.0293(0.99) &$0.030^{i_3}$&&&& \\
	5f8 $\rightarrow$ 5g10 &1.0246(0.99) &$1.035^{i_3}$&&&& \\
	6s2 $\rightarrow$ 6p2 &0.4778(1.13)&&0.4921&&& \\
	6s2 $\rightarrow$ 6p4 &1.0555(1.13)&&1.1260&&& \\
	6p2 $\rightarrow$ 6d4 &1.6248(1.11)&&1.5953&&& \\
	6p4 $\rightarrow$ 6d6 &1.4527(1.11)&&1.4249&&& \\
	6p4 $\rightarrow$ 6d4 &0.1601(1.11)&&0.1566&&& \\
	
	Rn$^{7+}$ \\
	6s2 $\rightarrow $ 6p2  &0.2372(1.00) &$0.234^{a_4}$&0.2379&&& \\
	6s2 $\rightarrow $ 6p4  &0.6880(1.02) &$0.689^{a_4}$&0.6922&&& \\
	6p2 $\rightarrow$ 7s2 &0.1634(1.06)  &$0.173^{a_4}$&&&& \\
	6p4 $\rightarrow$ 7s2 &0.2472(1.07)  &$0.259^{a_4}$&&&& \\ 
	6p2 $\rightarrow$6d4 &1.0154(1.01)   &$1.059^{a_4}$&1.0113&&& \\
	6p4 $\rightarrow$6d4 &0.1001(1.01)   &$0.103^{a_4}$&0.0999&&& \\
	6p4 $\rightarrow$6d6 &0.9234(1.01)   &$0.956^{a_4}$&0.9170&&& \\
	5f6 $\rightarrow$ 6d4 &0.0660(2.13)  &$0.0753^{a_4}$&0.0578&&& \\
	5f8 $\rightarrow$ 6d6 &0.0842(1.83)  &$0.0981^{a_4}$&0.0755&&& \\
	5f6 $\rightarrow$ 6d6 &0.0057(1.73)  &$0.0069^{a_4}$&0.0051&&&  \\
	5f6 $ \rightarrow$ 5g8 &0.6982(1.00)&&&&& \\
	7s2 $ \rightarrow$ 7p2 &0.4182(1.11) &$0.414^{a_4}$&&&& \\
    7s2 $ \rightarrow$ 7p4 &1.1727(1.12) &$1.126^{a_4}$&&&& \\
    6d4 $ \rightarrow$ 7p2 &0.2940(1.05) &$0.292^{a_4}$&&&& \\
    6d4 $ \rightarrow$ 7p4 &0.0513(1.04) &$0.052^{a_4}$&&&& \\
    6d6 $ \rightarrow$ 7p4 &0.3207(1.05) &$0.33^{a_4}$&&&& \\
	\hline	
\end{longtable}
\begin{flushleft}
        {${a_1}$$\implies$Beam-foil technique \cite{Reistad1986} \\
		$b_1$ $\implies$Calculations are based on high level methods such as the R-matrix method and asymptotic techniques developed by seaton \cite{Verner1996} \\
		$c_1$$\implies$ Single Configuration Interaction Hartree-Fock method using a pseudopotential \cite{Feret1999} \\
		$d_1$ $\implies$Nonrelativistic WKB approaches(Klein-Gordon dipole matrix) \cite{Lagmago1998} \\
		$e_1$ $\implies$Single configuration Dirac-Fock method \cite{Siegel1998} \\	
		$f_1$ $\implies$Relativistic many-body perturbation theory \cite{Guet1990} \\
		${g_1}$ $\implies$Realistic model potential \cite{Theodosiou1988} \\
		$h_1$ $\implies$Relativistic Hartree Fock method \cite{Kim1978} \\
		$i_1$ $\implies$multiconfiguration Dirac Hartee-Fock approximation.  \cite{Fischer2006} \\
		$j_1$ $\implies$Relativistic effective orbital quantum number \cite{Gruzdev1979} \\
		$a_2$ $\implies$Jointly analyzed decay curves: Beam-foil \cite{Livingston1980} \\
		$b_2$ $\implies$Multiexponential fits: Beam-foil \cite{Livingston1980} \\
		$c_2$ $\implies$Non-Relativistic Multi Configuration Hartree-Fock approximation \cite{Fischer1977} \\
		$d_2$ $\implies$ Relativistic Hartree-Fock \cite{Cheng1978} \\
		$e_2$ $\implies$Hartree-Fock oscillator strength using the Dirac correction factor \cite{Weiss1977} \\
		$f_2$ $\implies$Arbitrarily Normalized Decay curve method for cascade-correction in beam-foil  \cite{Pinnington1979} \\
		$g_2$ $\implies$Hartree-Fock with relativistic correction  \cite{Cowan1977} \\
		$h_2$ $\implies$Semi-empirical Coulomb approximation \cite{Lindgard1980} \\
		$i_2$ $\implies$model potential \cite{Migdalek1979} \\
		$j_2$ $\implies$Hartree-slater method \cite{Curtis1989} \\
		$a_3$ $\implies$RPTMP \cite{Ivanova2011} \\
		$b_3$ $\implies$RMBPT(3) \cite{Safronova2003} \\
		$c_3$ $\implies$DF+CP  \cite{Migdalek2000} \\
		$d_3$ $\implies$DX+CP method with SCE model potential \cite{Migdalek2000} \\
		$e_3$ $\implies$DX+CP method with CAFEGE model potential \cite{Migdalek2000} \\
		$f_3$ $\implies$DX+CP method with HFEGE model potential \cite{Migdalek2000} \\
		$g_3$ $\implies$CIDF method with integer occupation number \cite{Glowacki2009} \\
		$h_3$ $\implies$CIDF(q) method with non-occupation number \cite{Glowacki2009} \\
		$i_3$ $\implies$HFR+CP method \cite{Biemont2007} \\
		$a_4$$\implies$Relativistic core-polarization corrected Dirac-Fock method(DF+CP) \cite{Migdalek2018} \\
	}
\end{flushleft}

\LTcapwidth=\linewidth
\begin{longtable}{l l l l l l l l  }

 \caption {Transition rate (in $s^{-1}$) of E2 ($A_{RCC}^{E2}$) in plasma screened and unscreened medium.  Here we have used our RCC matrix element (in a.u) and RCC wavelength ( in \AA). Note that the notation $P(Q)$ in the case of transition rates means $P\times10^{Q}$.}\\ 
	\hline\hline
	Transitions     &$\mu$=0  & $\mu$=0.025 &$\mu$=0.050 &$\mu$=0.075&$\mu$=0.1 \\
	\hline
	\endfirsthead
	\multicolumn{7}{c}%
	{{\bfseries \tablename\ \thetable{} -- continued from previous page}} \\
	\hline
	
	Transition   & $\mu$=0 &$\mu$=0.025 &$\mu$=0.050 &$\mu$=0.075&$\mu$=0.1 \\ 
	\hline
	\endhead
	\hline
	\multicolumn{7}{l}%
		{Continued on next page}  \\
	\endfoot
	\hline
	\endlastfoot
	
	Ar$^{7+}$   & & & & &\\

	3p4 $\rightarrow$ 4f6  &1.4501(+06)  &1.4349(+06)&1.3927(+06)&1.3250(+06)&1.2333(+06) \\
    3p4 $\rightarrow$ 4f8  &6.5288(+06)  &6.4605(+06)&6.2705(+06)&5.9658(+06)&5.5526(+06)\\
    3s2$\rightarrow$ 3d4  &2.1840(+05)  &2.2032(+05)&2.2179(+05)&2.2417(+05)& 2.2742(+05) \\
	3s2$\longrightarrow$3d6&2.1919(+05)  &2.2126(+05)&2.2256(+05)&2.2494(+05)&2.2819(+05)\\
	3d4 $\rightarrow$ 5g8    &2.7225(+06)  &2.6258(+06)&2.3685(+06)&1.9874(+06)&\\
    3d6 $\rightarrow$ 5g8   &3.0252(+05)  & 2.9179(+05)&2.6352(+05)&2.2077(+05)& \\
	3d6 $\rightarrow$ 5g10&3.0267(+06)&2.9190(+06)&2.6362(+06)&2.2085(+06)&\\
	
	Kr$^{7+}$ & & & & & &  \\ 
	4s2$\longrightarrow $ 4d4 &8.9514(+05)&8.7868(+05)&8.6208(+05)&8.3473(+05)&\\
	4s2$\longrightarrow $ 4d6&9.1180(+05)&8.9640(+05)&8.7912(+05)&8.5066(+05)&\\	

	5p2 $\rightarrow$ 4f6  &7.0537(-01) &8.9779(-01)&3.0592(00)&&  \\
	5p4 $\rightarrow$ 4f8 &2.3536(-01) &2.9840(-01)&1.4702(00)&&   \\

	5d6 $\rightarrow$ 5g8 &2.7840(+03)&2.6100(+03)& &  &  \\
	5d6 $\rightarrow$ 5g10   &2.9352(+04) &2.6114(+04)& &&   \\
	
	&&&&& \\
	Xe$^{7+}$ & & & & & &  \\
	5s2$\rightarrow$ 5d4  &6.6714(+05) &6.7116(+05)&& & \\
	5s2$\rightarrow$ 5d6&6.9674(+05)&7.0069(+05) & \\	
	5p2 $\rightarrow$ 5p4  & 6.0045(-01)  &6.1555(-01)&5.6058(-01) \\
	5p2 $\rightarrow$ 4f6 &8.5467 (+03) &2.7288(+04)&3.3098(+04) \\
	5p4 $\rightarrow$ 4f6&1.2882 (+03) &2.0657(+03)&2.6049(+03) \\
	5p4 $\rightarrow$ 4f8  &5.9866 (+03) &1.2836(+04)&1.6523(+04)   \\
	5p2$\rightarrow$ 5f6&1.6471(+06) &1.5379(+06) &  \\
	5p4$\rightarrow$ 5f6&4.3300(+05) &4.0395(+05) &  \\
	5p4$\rightarrow$ 5f8  &1.9479(+06)  &1.8169(+06) &   \\
	4f6 $\rightarrow$ 5f6  &5.9296 (+04)  &5.4103(+04) &  \\
	4f6 $\rightarrow$ 5f8   &7.4168 (+03)  &6.7680(+03) &  \\
	4f8 $\rightarrow$ 5f6  &9.9224 (+03)  &9.0350(+03) &  \\
	4f8 $\rightarrow$ 5f8   &6.2129(+04)  & 5.6576(+04)& \\

	5d4$\rightarrow$ 6s2&5.3813(+03) & &  \\
	5d4 $\rightarrow$ 5g8 &1.1670 (+06) & &   \\
	5d6 $\rightarrow$ 5g8&1.2597 (+05) & &  \\
	5d6$\rightarrow$ 5g10&1.2601(+06) & &   \\
	5d6$\rightarrow$ 6s2  &7.0259 (+03) & &   \\
	Rn$^{7+}$ & & & & & &  \\
	6s2 $\rightarrow$ 6d4     &1.0243(+06)   &9.8377(+05)&&& \\
	6s2 $\rightarrow$ 6d6  &1.0371(+06)&9.9749(+05)&&& \\	
	6p2 $\rightarrow$5f6   &5.0885(+04) &5.4520(+04)&& &  \\
	6p4 $\rightarrow$5f6&3.2019(+03) &3.5290(+03) &&& \\
	6p4 $\rightarrow$5f8&1.5773(+04) &1.7127(+04) &&& \\
	6d4 $\rightarrow$ 5g8&6.3782(+05)&&  & &  \\
	6d6 $\rightarrow$ 5g8   &6.2707(+04)   &&& &  \\
	6d6 $\rightarrow$ 5g10   &6.2238 (+05)&& & &  \\
	6p2 $\rightarrow$ 6p4    &1.2855 (+02)  &1.2763(+02)&& &  \\

    7p2 $\rightarrow$ 7p4    &1.5299 (+01)  &&& &  \\
    6p4 $\rightarrow$ 7p4    &2.3245 (+05)  &&& &  \\
    5f6 $\rightarrow$ 7p2    &1.4611 (+04)  &&& &  \\
    5f6 $\rightarrow$ 7p4    &3.0317 (+03)  &&& &  \\
    5f8 $\rightarrow$ 7p4    &1.7756 (+04)  &&& &  \\
	\hline	
\end{longtable}

\LTcapwidth=\linewidth
\begin{longtable}{l l l l l l l l }
		\caption {Magmnetic dipole transition rate (in $sec^{-1}$) in plasma screened and unscreened medium. Notes: The notation $P(Q)$ in the case of transition rates means $P\times10^{Q}$. For $4f_{5/2}\ \rightarrow\ 4f_{7/2}$ (Xe$^{7+}$),  transition rates 1.9227 (-03) and 1.9277 (-03) are available in the literature (a) using multiconfiguration Dirac-Fock method without and with Breit interaction plus quantum electrodynamics effect, respectively .} \\
			
	\hline\hline
	Transition  & $\mu$=0 &$\mu$=0.025&$\mu$=0.050&$\mu$=0.075&$\mu$=0.1  \\
\hline
\endfirsthead

Transition   & $\mu$=0 &$\mu$=0.025 &$\mu$=0.050 &$\mu$=0.075&$\mu$=0.1 \\ 
\hline
\endhead

\endfoot

\endlastfoot

	Ar$^{7+}$  \\

	3p2 $\rightarrow$ 3p4    & 1.7951(-01) &1.9857(-01)&1.9603(-01)&1.9165(-01)&1.8570(-01)\\
	3p4 $\rightarrow$ 4f6    & 3.7845(-02) &3.7248(-02)&3.5575(-02)&3.2958(-02)& 2.9561(-02) \\

	3d6 $\rightarrow$ 5g8    &1.0634 (-02) &1.0124(-02)&8.8719(-03)&7.0637(-03)& \\
	Kr$^{7+}$   \\ 
	4s2$\rightarrow $ 4d4   & 9.0675(-03) & 7.2173(-03)&8.2040(-03)&6.7390(-03)&       \\	
	4p2 $\rightarrow$ 4p4    &8.4015(00)   &8.2789(00)  &7.9664(00) &7.4776(00) &6.8397(00) \\	

	4d4 $\rightarrow$ 4d6    &2.5436(-02)  &3.2234(-02) &3.0177(-02)&2.7000(-02)&        \\
	5p2 $\rightarrow$ 5p4    &4.7841(-01)  &5.3517(-01) &4.6911(-01)&           &        \\
    
	5d4 $\rightarrow$ 5d6  &1.7080(-03)  &1.0904(-03) &2.2682(-03)&           &        \\
	Xe$^{7+}$               \\
	
	5p2 $\rightarrow$ 5p4    &5.7437(+01)  &5.7875(+01) &5.3500(+01)  &           & \\

	5p4 $\rightarrow$ 5f6    &6.3588(-03)  &5.7557(-03) &           &           &  \\
	
	4f6 $\rightarrow$ 5f6    &5.0459(+01)  &4.9605(+01) &           &           &\\
	4f6 $\rightarrow$ 5f8   &4.9212(-01)  &5.3389(-01) &           &           &\\
	4f8 $\rightarrow$ 5f6   &1.1142(+01)  &1.0620(+01) &           &           &\\
	4f8 $\rightarrow$ 5f8   &1.5833 (+02) &1.5517(+02) &           &           & \\
	5d4 $\rightarrow$ 5d6    &2.7070(-01)  &2.6538(-01) &2.2900(-01)  &           & \\

	Rn$^{7+}$  \\

	6p2 $\rightarrow$ 6p4    &1.2442 (+03) &1.2212(+03) &           &           & \\
	6d4 $\rightarrow$ 6d6    &4.5873(00)   &4.1847(00) &            &           & \\
	5f6 $\rightarrow$ 5f8    &6.6129 (-02) &3.4422(-02) &           &           & \\

	6s2 $\rightarrow$ 7s2    &1.0113 (00)  &            &           &           &   \\
  7p2 $\rightarrow$ 7p4    &8.1938 (+01)  &            &           &           &   \\
  6p2 $\rightarrow$ 7p2    &7.2065 (-02)  &            &           &           &   \\
  6p4 $\rightarrow$ 7p4    &2.4557 (00)  &            &           &           &   \\
  5f6 $\rightarrow$ 7p4    &2.4281 (-03)  &            &           &           &   \\

	\hline
	\multicolumn{7}{l}{\textsuperscript{a}\footnotesize{\cite{Grumer2014, Ding2012}}}	
\end{longtable}

\LTcapwidth=\linewidth
\begin{longtable}{c c c c }
\caption{Lifetimes  in ns of few low-lying states. }\\
 
Level &present work &other work(experiment) &other work(theory) \\
\hline
 Ar$^{7+}$& & & \\
3p2 &0.411 & 0.417
$\pm0.010^a$, 0.423 $\pm0.040^b$, & $0.413^c$, $0.407^d$, $0.397^e$, $0.409^f$ \\
& & 0.48$\pm0.05^g$, 0.49$\pm0.05^{h}$&$0.389^{i}$, $0.408^{j}$, $0.4121^{k}$ \\
& & 0.55$\pm0.03^{l}$, 0.53$\pm0.11^{m}$ & \\
3p4 &0.387 & 0.389$ \pm0.010^a $, 0.421$\pm0.030^b$,  
&$0.389^c$, $0.382^d$, $0.373^e$, $0.386^f$ \\
& & 0.428$\pm0.027^g$, 0.48$\pm0.06^h$&$0.366^i$, $0.388^j$, $0.3872^{k}$ \\
& &0.54$\pm0.0.02^{l}$, 0.527$\pm0.018^{m}$& \\
3d4 &0.132 & 0.170
$\pm0.0.010^a$, 0.130 $\pm0.005^b$, &$0.127^e$, $0.134^f$, $0.133^j$,  $0.1318^{k}$ \\
& & 0.158$\pm0.008^g$, & \\
3d6 &0.137 & 0.166
$\pm0.008^a$, 0.131 $\pm0.005^b$, &$0.131^e$, $0.138^{f, j}$,   $0.1361^{k}$ \\
& &0.160$\pm0.008^g$ \\
4f6 &0.003 & & \\
4f8&0.002 & & \\

 Kr$^{7+}$& & & \\
4p2 &0.293 & 0.41
$\pm0.04^{n}$, $0.291\pm0.012^{o}$ &$0.282^{p}$, $0.29653^{q}$ \\
&&$0.290\pm0.015^g$, $0.401\pm0.018^{l}$& \\
4p4 & 0.235& 0.33$ \pm0.03^{n}$, $0.243\pm0.01^{o}$ &$0.230^{p}$, $0.24176^{q}$ \\
&&$0.218\pm0.033^g$, $0.331\pm0.011^{l}$\\
4d4 &0.048 &  &$0.05019^{q}$ \\
4d6 &0.052 &$0.048\pm0.004^{o}$&$0.05388^{q}$  \\
4f6&0.055&&\\
4f8&0.055&& \\

Xe$^{7+}$ & & & \\
5p2 &0.45 & $0.52(3)^{r}$,
0.50 $\pm0.05^{s}$, & $0.37^{t}$, $0.47^{u}$, $0.48^{v}$, $0.53^{w}$ \\
&&$0.380\pm0.040^g$\\
5p4 &0.29 & $0.35(2)^{r}$, 0.33$\pm0.03^{s}$,  
&$0.23^{t}$, $0.30^{u}$, $0.31^{v}$, $0.33^{w}$ \\
&& $0.272\pm0.037^g$\\
5d4 &0.08 &$0.10(2)^{r}$  &$0.07^{t}$, $0.07^{v}$, $0.06^{w}$ \\
5d6 &0.08 & $0.14(2)^{r}$ &$0.14^{t}$, $0.08^{v}$, $0.07^{w}$ \\
 Rn$^{7+}$& & & \\
6p2 &0.429 & &  \\
6p4 &0.144 & & \\
6d4 &0.056 &  & \\
6d6 &0.082 &  & \\
\hline \\

\end{longtable}
\begin{flushleft}
{$a$$\implies$Beam-Foil technique \cite{Reistad1986} \\
$b$$\implies$\cite{Buchet-Poulizac1982} \\
$c$$\implies$Third order many-Body pertrubation theory \cite{Johnson1996} \\
$d$$\implies$  R-matrix theory \cite{Verner1996} \\
$e$$\implies$Single Configuration interaction Hartree-Fock method using a pseudo potential \cite{Feret1999} \\
$f$$\implies$ Relastic model potential \cite{Theodosiou1988} \\
$g$$\implies$Arbitrarily Normalized Decay curve method for cascade-correction in beam-foil \cite{Pinnington1979} \\
$h$$\implies$Beam-Foil technique in the vacuum u.v \cite{Knystautas1979} \\
$i$$\implies$Multiconfiguration Dirac-Fock method \cite{Kim1978} \\
$j$$\implies$Charge expansion technique \cite{Crossley1965} \\
$k$$\implies$Multiconfiguration Dirac Hartree-Fock theory including core polarisation \cite{Fischer2006} \\
$l$$\implies$Beam-Foil \cite{Irwin1976} \\
$m$$\implies$Beam-Foil \cite{Livingston1972} \\
$n$$\implies$Beam-Foil \cite{Druetta1976} \\
$o$$\implies$Foil excitation \cite{Livingston1980} \\
$p$$\implies$Coulomb approximation \cite{Lindgard1980} \\
$q$$\implies$Hartree-Slater method \cite{Curtis1989} \\
$r$$\implies$Beam-foil spectroscopy \cite{Biemont2007} \\
$s$$\implies$Relativistic Hartree-Fock method \cite{Cheng and Kim1979} \\
$t$$\implies$Relativistic perturbation theory with a zero approximation model potential \cite{Ivanova2011} \\
$u$$\implies$Relativistic Many-Body perturbation theory(RMBPT(3)) \cite{Safronova2003} \\
$v$$\implies$Relativistic HFR+CP \cite{Biemont2007} \\
$w$$\implies$Relativistic MCDF \cite{Biemont2007} \\
}
\end{flushleft}

\begin{table*}
	\caption{Critical values of plasma screening strength ($\mu_{c}$) in a.u for the following ions.  }
 \centering
	\begin{tabular}{cccccc}
		\hline\hline
	\multicolumn{2}{c}{Kr$^{7+}$}&\multicolumn{2}{c}{Xe$^{7+}$}&\multicolumn{2}{c}{Rn$^{7+}$}\\
		state &$\mu_{c}$&state &$\mu_{c}$&state &$\mu_{c}$\\
		\hline
		
	4d4 &0.087152& 5s2 &0.076987 &5s2 &0.046401\\
		4d6 &0.086956&5p2 &0.066149&6p2 &0.039652 \\
		&&5p4 &0.064474& 6p4 &0.036763 \\
	   && 4f6 &0.051526&6d4 &0.027521  \\
		&& 4f8 &0.051486& 6d6 &0.027100 \\
		&&5d4 &0.048395&& \\
		&&5d6 &0.048139&& \\
  \hline
	\end{tabular}
 
\end{table*}

	\begin{figure}
	\caption{Determination of the value of critical plasma screening strength }
	\begin{subfigure}{0.5\textwidth}
		\centering
		\includegraphics[width=0.8\linewidth]{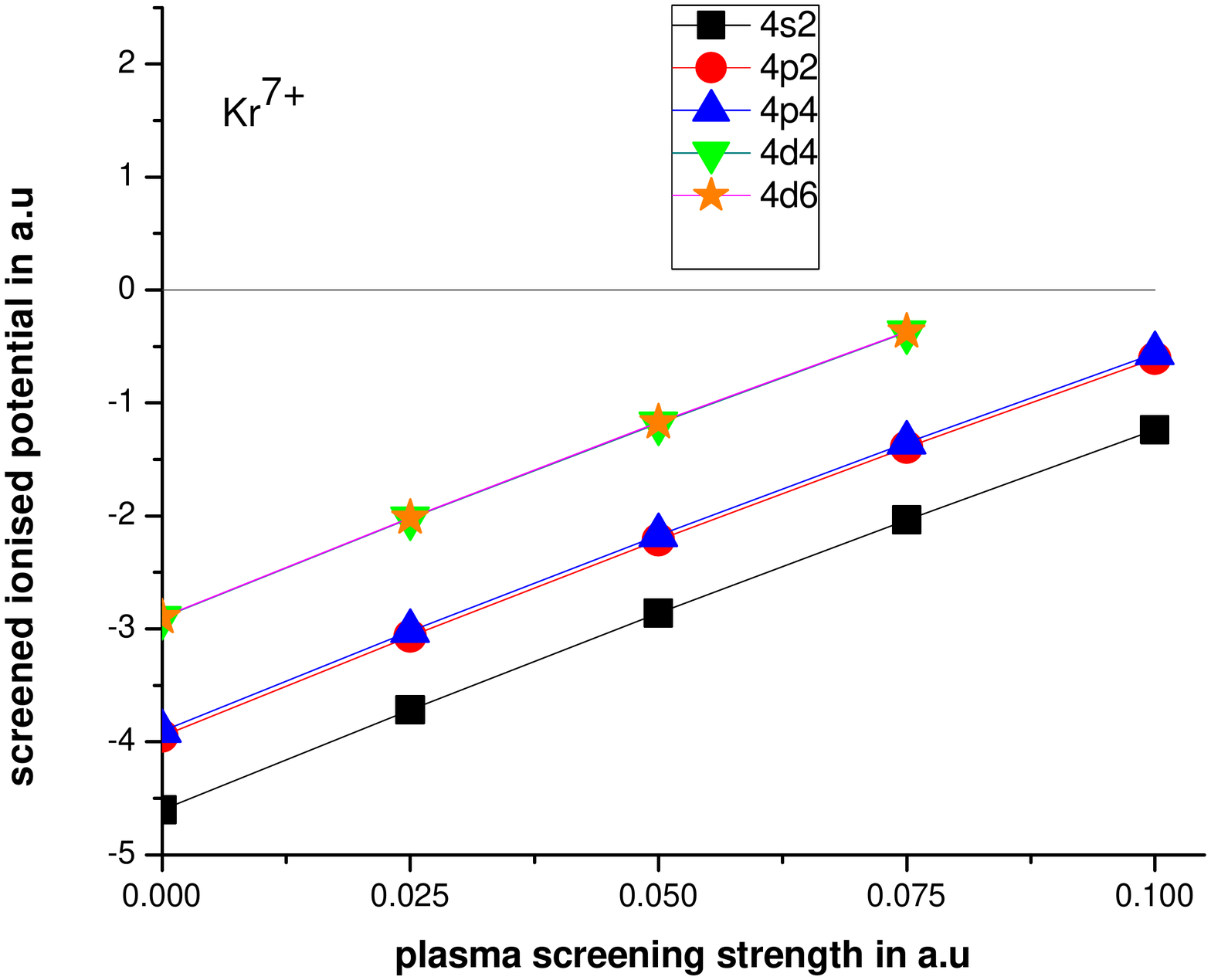}

	\end{subfigure}
	\begin{subfigure}{0.5\textwidth}
		\centering
		\includegraphics[width=0.8\linewidth]{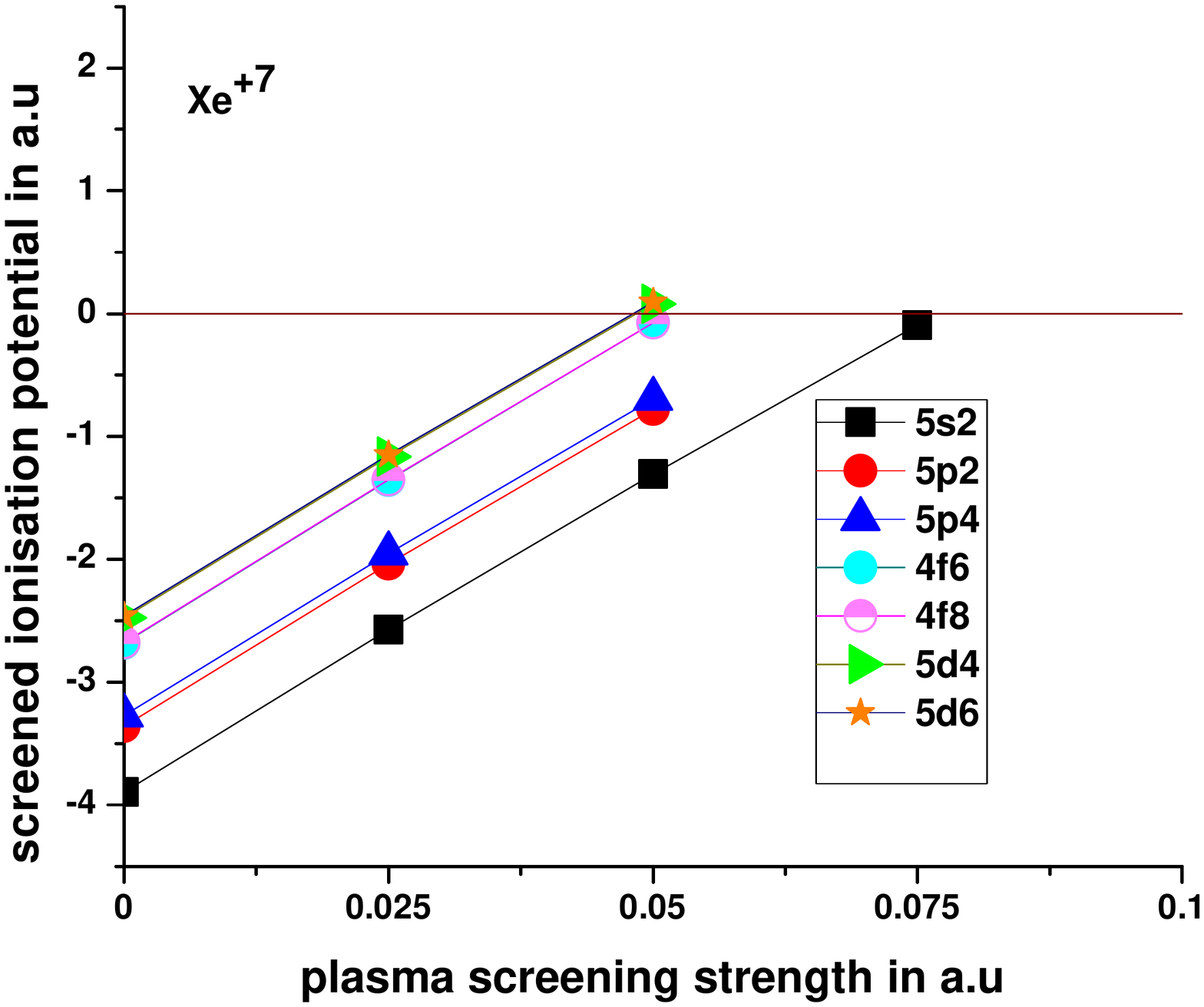}

	\end{subfigure}
	\begin{subfigure}{0.5\textwidth}
		\centering
		\includegraphics[width=0.8\linewidth]{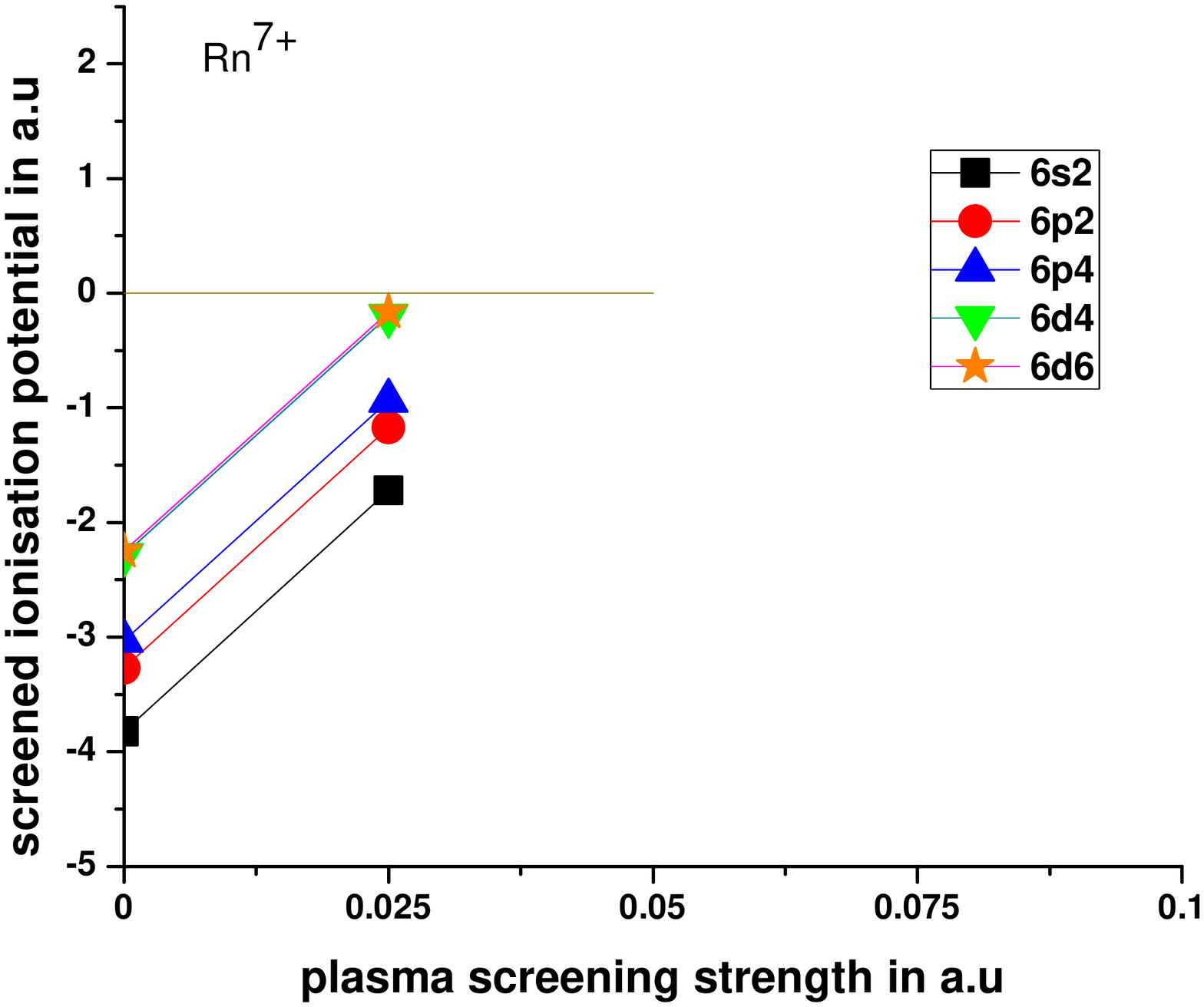}

	\end{subfigure}
\end{figure}

\end{document}